\documentclass[prd,nofootinbib]{revtex4}
\usepackage{epsfig,amsmath,slashed}
\newcommand{\bra}[1]{\langle #1|}
\newcommand{\ket}[1]{|#1\rangle}

\newcommand{\di}{{\rm d}}
\newcommand{\ii}{i}
\newcommand{\Pro}{{\sf P}}
\def\spt{{\cal S}}
\def\wT{{\widehat T}}
\def\wspt{{\widehat{\cal S}}}

\def\wPhi{{\widehat{\Phi}}}
\def\wpsi{{\widehat{\psi}}}
\def\wrho{{\widehat{\rho}}}

\def\codev{{\stackrel{\leftrightarrow}{\partial}}}
\newcommand{\tr}{{\rm tr}}  
  
\newcommand{\e}{{\rm e}}

\newcommand{\omegav}{\boldsymbol{\omega}}

\newcommand{\x}{{\rm x}}

\newcommand{\Psibar}{{\overline \Psi}}

\newcommand{\zet}{\zeta_{(M,\xi,l)}}
\newcommand{\ppl}{p_{+,\xi}}
\newcommand{\pmi}{p_{-,\xi}}
\newcommand{\pmm}{p_{-,-\xi}}
\newcommand{\pt}[1]{{\left( #1 \right)}}                                              
\newcommand{\pq}[1]{{\left[ #1 \right]}}                                              
                                  
\newcommand{\n}{{\bf n}}                                                            
\newcommand{\abs}[1]{\left| #1 \right|}                                             
\newcommand{\besselbetauno}[1]{J_{\abs{M-\frac{1}{2}}} \pt{p_{Tl} #1} }             
\newcommand{\besselbetadue}[1]{J_{\abs{M+\frac{1}{2}}} \pt{p_{Tl} #1} }             
\newcommand{\besselquadrouno}[1]{J_{\abs{M-\frac{1}{2}}}^2 \pt{p_{Tl} #1}}          
\newcommand{\besselquadrodue}[1]{J_{\abs{M+\frac{1}{2}}}^2 \pt{p_{Tl} #1}}          
\newcommand{\bsbu}[1]{J_{\abs{M-\frac{1}{2}}}\pt{#1}}                               
\newcommand{\bsbd}[1]{J_{\abs{M+\frac{1}{2}}}\pt{#1}}                               
\newcommand{\sgnm}{{\rm sgn}(M)}                                              
\newcommand{\ks}{\kappa_\xi}                                                       
\newcommand{\bpms}[2]{b_{#1 \xi}^{\pt{#2}} }                                        
\newcommand{\esp}[1]{{\rm e}^{#1} }

\begin{document}

\title{Thermodynamical inequivalence of quantum stress-energy and spin tensors} 

\author{F. Becattini, L. Tinti}\affiliation{Universit\`a di 
 Firenze and INFN Sezione di Firenze, Florence, Italy} 

\begin{abstract}
It is shown that different couples of stress-energy and spin tensors of 
quantum relativistic fields, which would be otherwise equivalent, are in fact 
inequivalent if the second law of thermodynamics is taken into account. The proof 
of the inequivalence is based on the analysis of a macroscopic system at full 
thermodynamical equilibrium with a macroscopic total angular momentum and a 
specific instance is given for the free Dirac field, for which we show that the 
canonical and Belinfante stress-energy tensors are not equivalent. For this 
particular case, we show that the difference between the predicted angular 
momentum densities for a rotating system at full thermodynamical equilibrium 
is a quantum effect, persisting in the non-relativistic limit, corresponding
to a polarization of particles of the order of $\hbar \omega/KT$ ($\omega$ 
being the angular velocity) and could in principle be measured experimentally.
This result implies that specific stress-energy and spin tensors are physically 
meaningful even in the absence of gravitational coupling and raises the issue of 
finding the thermodynamically right (or the right class of) tensors. We argue that 
the maximization of the thermodynamic potential theoretically allows to discriminate 
between two different couples, yet for the present we are unable to provide a
theoretical method to single out the ``best" couple of tensors in a given quantum 
field theory. The existence of a non-vanishing spin tensor would have major 
consequences in hydrodynamics, gravity and cosmology.
\end{abstract}

\maketitle

\section{Introduction}
\label{intro}

It is commonly known that stress-energy and spin tensors are not uniquely defined 
in field theory as long as gravity is disregarded. In quantum field theory, distinct 
stress-energy tensors differing by the divergence of a rank 3 tensor provide, once 
integrated in three-dimensional space, the same generators of space-time translations
provided that the flux of the additional rank 3 tensor field (hereafter referred to as
superpotential) vanishes at the boundary. Correspondingly, in classical field theory, 
the spatial integrals of $T^{0 \nu}$ yield the same values of total energy and momentum.
Within field theory on a flat spacetime, it is then possible to generate apparently 
equivalent stress-energy tensors which are e.g. symmetric or non-symmetric. Indeed, 
gravitational coupling provides an unambiguous way of defining the stress-energy tensor; 
in General Relativity, it is symmetric by construction and the spin tensor vanishes. 
However, in a likely extension known as Einstein-Cartan theory (not excluded by present
observations) the spin tensor is non-vanishing and the stress-energy tensor is 
non-symmetric.

Can we say something more? In classical physics we have a stronger requirement with 
respect to a quantum theory: we would like the energy, momentum and angular momentum 
content of {\em any} arbitrary macroscopic spatial region to be well defined concepts; 
otherwise stated, we would like to have objective values for the energy, momentum and 
angular momentum densities. If these quantities are to be the components of the 
stress-energy and spin tensors, such a requirement strongly limits the freedom to 
change these tensors. It is crucial to emphasize, from the very beginning, the difference 
between quantum and classical tensors. The quantum stress-energy and spin tensors, 
henceforth denoted with a hat $\widehat {\phantom a}$, are operatorial expressions 
depending on the microscopic quantum field operators $\Psi$, whereas the classical 
ones are c-numbers. The relation between them is \cite{degroot}:
\begin{equation}\label{mean}
 T^{\mu \nu}(x) = \tr [ \widehat \rho \, :\!\wT^{\mu \nu}(x)\!: ] \qquad
 \spt^{\lambda,\mu \nu}(x) = \tr [ \widehat \rho \, :\!\wspt^{\lambda, \mu \nu}(x)\!:] 
\end{equation}
where $\widehat \rho$ is the density operator describing the (mixed or pure) quantum 
state and $:$ denotes normal ordering; the latter is usually introduced in the mean
value definition in order to subtract the zero-point infinities \footnote{For a 
discussion of the meaning of normal ordering for interacting fields see e.g. 
ref.~\cite{deweert}. We stress that the results obtained in this work, particularly 
in Sect.~\ref{dirac} are anyhow independent of the use of normal ordering in eq.~(\ref{mean}).}. 
According to (\ref{mean}), a change of quantum stress-energy and spin tensors could 
induce a change of the corresponding classical ones in an undesirable fashion, meaning
that energy or momentum or angular momentum density get changed. However, the change 
that classical mean values undergo as a reflection of a variation of quantum tensors 
crucially depends on the physical state $\wrho$. Particularly, we will see that the 
freedom of varying the stress-energy and spin tensors at a quantum level depends on 
the symmetry features of the physical state: a highly symmetric state allows more 
changes of quantum tensors than a state with little symmetry does.

In this paper, we prove that a system at full thermodynamical equilibrium with
a macroscopic value of angular momentum, thence rigidly rotating \cite{landau}, 
allows to discriminate between different quantum spin tensors, and, consequently, 
between different quantum stress-energy tensors. This kind of inequivalence shows 
up only for a rotating system whereas all quantum tensors are equivalent for a 
system at the more familiar thermodynamical equilibrium with vanishing macroscopic 
angular momentum.
The paper is organized as follows: in Sect.~\ref{transform} we will discuss the 
general class of stress-energy tensor transformations ensuring the invariance 
of conservation equations; in Sect.~\ref{thermo} we will discuss the usual 
thermodynamical equilibrium distribution and its symmetries and in Sect.~\ref{thermang} 
we will do the same for a system at full thermodynamical equilibrium with angular 
momentum; in Sect.~\ref{axisymm} we will obtain the most general form of mean
stress-energy and spin tensor for a system at full thermodynamical equilibrium with 
angular momentum and show that the equivalence between different quantum 
tensors no longer applies unless peculiar conditions are met; in Sect.~\ref{dirac} we 
will present and prove a concrete instance of inequivalence for the free Dirac field; 
finally, in Sect.~\ref{concl} we will summarize and further illustrate the obtained 
result and discuss the possible consequences thereof.

\subsection*{Notation}

In this paper we adopt the natural units, with $\hbar=c=K=1$.\\
The Minkowskian metric tensor is ${\rm diag}(1,-1,-1,-1)$; for the Levi-Civita
symbol we use the convention $\epsilon^{0123}=1$.\\ 
We will use the relativistic notation with repeated greek indices assumed to 
be saturated. Operators in Hilbert space will be denoted by an upper hat, e.g. 
$\widehat {\sf R}$, with the exception of the Dirac field operator which is 
denoted with a capital $\Psi$.

\section{Transformations of stress-energy and spin tensors}
\label{transform}

The conservation equations ensuing from the relativistic translational and Lorentz
invariance are the well known continuity equations of energy-momentum and total
angular momentum: 
\begin{eqnarray}\label{conservq}
 && \partial_\mu \wT^{\mu \nu} = 0 \nonumber \\
 && \partial_\lambda \widehat{\cal J}^{\lambda, \mu \nu} = \partial_\lambda 
 \left ( \wspt^{\lambda, \mu \nu} + x^\mu \wT^{\lambda \nu} - x^\nu \wT^{\lambda \mu} 
 \right) = \partial_\lambda \wspt^{\lambda, \mu \nu} + \wT^{\mu \nu} - \wT^{\nu \mu} 
 = 0
\end{eqnarray}
However, stress-energy $\wT$ and spin tensor $\wspt$ are not uniquely defined in 
quantum field theory; once a particular couple $(\wT,\wspt)$ of these tensors is 
found, e.g. applying Noether's theorem to some Lagrangian density (the so-called 
{\em canonical} tensors), it is possible to generate new couples $(\wT',\wspt')$ 
through the following pseudo-gauge transformation \cite{halbw}:
\begin{eqnarray}\label{transfq}
 && \wT'^{\mu \nu} = \wT^{\mu \nu} +\frac{1}{2} \partial_\alpha
 \left( \wPhi^{\alpha, \mu \nu } - \wPhi^{\mu, \alpha \nu} - 
 \wPhi^{\nu, \alpha \mu}  \right) \nonumber \\
 && \wspt'^{\lambda, \mu \nu} = \wspt^{\lambda,\mu\nu}-\wPhi^{\lambda,\mu\nu}
\end{eqnarray}
where $\wPhi$ is an arbitrary tensor of rank three antisymmetric in the last two 
indices depending on the fields $\wpsi$. It is easy to check that the new couple 
fulfills the same continuity equations (\ref{conservq}) as the original one and
that the new total angular momentum tensor, like the stress-energy tensor, differs
from the original one by a divergence:
\begin{equation}\label{transfq2}
 \widehat{\cal J}'^{\lambda, \mu \nu} = \widehat{\cal J}^{\lambda, \mu \nu}
 + \frac{1}{2} \partial_\alpha \left[ x^\mu \left( \wPhi^{\alpha, \lambda \nu } - 
 \wPhi^{\lambda, \alpha \nu} - \wPhi^{\nu, \alpha \lambda} \right) - x^\nu \left( 
 \wPhi^{\alpha, \lambda \mu } - \wPhi^{\lambda, \alpha \mu} - 
 \wPhi^{\mu, \alpha \lambda} \right) \right]
\end{equation}

The spatial integrals over the domain $\Omega$:
\begin{eqnarray}\label{generators}
 && \widehat P^\nu = \int_\Omega \di^3 \x \;  \wT^{0 \nu} \nonumber \\
 && \widehat J^{\mu \nu}=  \int_\Omega \di^3 \x \;  \widehat{\cal J}^{0, \mu \nu} = 
  \int \di^3 \x \; \wspt^{0, \mu \nu} + x^\mu \wT^{0 \nu} - x^\nu \wT^{0\mu}
\end{eqnarray}
are conserved (and are generators of translations and Lorentz transformations if
the domain is the whole space) provided that the fluxes at the boundary vanish:
\begin{eqnarray}\label{flux}
 &&  \int_{\partial \Omega} \di S \;  \wT^{i \nu} n_i = 0 \nonumber \\
 &&  \int_{\partial \Omega} \di S \; (\wspt^{i, \mu \nu} + 
  x^\mu \wT^{i \nu} - x^\nu \wT^{i \nu}) n_i = 0
\end{eqnarray}  
$n_i$ being the surface normal versor. The eqs.~(\ref{flux}) are usually ensured 
by enforcing special boundary conditions for the fields; for instance, the familiar 
periodic boundary conditions for a box, or requiring some expression involving 
the field or its normal derivatives to vanish at the boundary (this will be the 
case in this paper, see Sect.~\ref{dirac}). It is very important to stress here 
that one can obtain conserved stress-energy and angular momentum tensors and
a corresponding set of conserved generators even within a finite or bounded region,
thus breaking global translational and rotational symmetry, provided that 
suitable boundary conditions are imposed on the fields; this point will be further 
discussed in Sect.~\ref{dirac}. 

The spatial integrals (\ref{generators}) of the new tensors are invariant, thus 
yielding the same generators, if the tensor $\wPhi$ is such that the following 
boundary integrals vanish, according to eq.~(\ref{transfq}) and (\ref{transfq2}): 
\begin{eqnarray}\label{boundary}
 && \int_{\partial \Omega} \di S \; \left( \wPhi^{i, 0 \nu } - \wPhi^{0, i \nu} - 
 \wPhi^{\nu, i 0} \right) n_i = 0 \nonumber \\
 && \int_{\partial \Omega} \di S \; \left[ x^\mu \left( \wPhi^{i, 0 \nu } - \wPhi^{0, i \nu} - 
 \wPhi^{\nu, i 0} \right) - x^\nu \left( \wPhi^{i, 0 \mu } - \wPhi^{0, i \mu} - 
 \wPhi^{\mu, i 0} \right) \right] n_i = 0 
\end{eqnarray}
In fact, if the above conditions are met, the flux integrals (\ref{flux}) of the 
new primed tensors vanish because the new tensors also fulfill the continuity equations. 
In conclusion, a pseudo-gauge transformation like (\ref{transfq}) is always possible 
provided that the boundary conditions (\ref{boundary}) are ensured; in this case 
the couple $(\wT,\wspt)$ and $(\wT',\wspt')$ are regarded as equivalent in quantum 
field theory because they give the same total energy, momentum and angular momentum, 
in the operatorial sense.
 
The classical counterpart of transformation (\ref{transfq}) can be calculated by 
applying eq.~(\ref{mean}) to both sides and this obviously leads to:
\begin{eqnarray}\label{transf}
 && T'^{\mu \nu} = T^{\mu \nu} +\frac{1}{2} \partial_\alpha
 \left( \Phi^{\alpha, \mu \nu } - \Phi^{\mu, \alpha \nu} - 
 \Phi^{\nu, \alpha \mu}  \right) \nonumber \\
 && \spt'^{\lambda, \mu \nu} = {\cal S}^{\lambda, \mu \nu} -
 \Phi^{\lambda,\mu \nu}
\end{eqnarray}
If the system is macroscopic, we would like the mean values of those tensors to
be invariant under a transformation (\ref{transf}), and not just their integrals.
This is because energy, momentum and total angular momentum densities classically
must take on objective values, independent of the particular quantum tensors.
A minimal requirement would be the invariance of the aforementioned densities,
that is:
\begin{equation*}
 T'^{0 \nu} = T^{0 \nu} \qquad {\cal J}'^{0, \mu \nu} = {\cal J}^{0,\mu \nu}
\end{equation*}
However, this is a frame-dependent requirement; a Lorenzt-boosted frame would 
measure a different energy-momentum density if only the first row of the 
stress-energy tensor was invariant under transformation (\ref{transf}) in one 
particular frame. We are thus to enforce a stricter requirement, namely:
\begin{equation}\label{req1}
 T'^{\mu \nu} = T^{\mu \nu}
\end{equation}
whereas, for the rank 3 angular momentum tensor, we can make a looser request:
\begin{equation}\label{req2}
 {\cal J}'^{\lambda,\mu \nu} = {\cal J}^{\lambda,\mu \nu} + 
 g^{\lambda \mu} K^\nu - g^{\lambda \nu} K^\mu  
\end{equation}
where $K$ is a vector field. Indeed, if we limit ourselves to spatial indices 
$\mu, \nu = 1,2,3$, the above equation is enough to ensure that the angular momentum 
densities, with $\lambda=0$, are the same in {\em any} inertial frame. Comparing 
eq.~(\ref{req1}) with eq.~(\ref{transf}), we get:
\begin{equation}\label{phieq}
 \partial_\alpha \left( \Phi^{\alpha,\mu \nu} - \Phi^{\mu,\alpha \nu} - 
 \Phi^{\nu, \alpha \mu}\right) = 0
\end{equation}
while comparing eq.~(\ref{req2}) with the mean of eq.~(\ref{transfq2}) and 
taking (\ref{phieq}) into account, we obtain a simple condition for the 
superpotential to meet:
\begin{eqnarray}\label{vanish}
 && \frac{1}{2} \partial_\alpha \left[ x^\mu \left( \Phi^{\alpha,\lambda \nu} - 
 \Phi^{\lambda, \alpha \nu} - \Phi^{\nu, \alpha \lambda} \right) - x^\nu 
 \left( \Phi^{\alpha, \lambda \mu} - \Phi^{\lambda,\alpha \mu} - 
 \Phi^{\mu,\alpha \lambda} \right) \right] =  
 g^{\lambda \mu} K^\nu - g^{\lambda \nu} K^\mu
 \nonumber \\
 \Rightarrow && \frac{1}{2} \left( 
 \Phi^{\mu,\lambda \nu} - \Phi^{\lambda, \mu \nu} - \Phi^{\nu, \mu \lambda} 
 - \Phi^{\nu, \lambda \mu} + \Phi^{\lambda,\nu \mu} + \Phi^{\mu,\nu \lambda} 
 \right) = - \Phi^{\lambda,\mu \nu} = g^{\lambda \mu} K^\nu - g^{\lambda \nu} K^\mu  
\end{eqnarray}
Plugging this last result back into eq.~(\ref{phieq}) one obtains:
\begin{equation*}
 2 \partial_\alpha (K^\mu g^{\alpha \nu} - K^\alpha g^{\mu\nu}) = 
 2 \partial^\nu K^\mu - 2 g^{\mu \nu} \partial \cdot K = 0
\end{equation*}
Contracting the indices $\mu$ and $\nu$ we obtain at once that the divergence 
of the vector field $K$ vanishes and so, because of the above equation: 
\begin{equation}\label{fincond}
  \partial^\nu K^\mu = 0
\end{equation}  
Therefore the eqs.~(\ref{phieq}) and (\ref{vanish}) imply that the vector field 
$K$ is a constant field. The possible directions of this field will be dictated
by the symmetry properties of the system under consideration, as we will see in the
next two sections.

It should be emphasized that the conditions (\ref{phieq}) and (\ref{vanish}) do
not need to apply to the quantum tensor $\wPhi$, which only has to meet
the boundary conditions (\ref{boundary}), as has been seen. On the other hand,
if we take the mean values of (\ref{boundary}) applying $\tr (\wrho \;\;)$ on 
both sides, the ensuing equation is a trivial consequence of the eq.~(\ref{phieq}). 
In fact, it may happen that the mean value of the superpotential $\Phi$ fulfills eqs.
(\ref{phieq}) and (\ref{vanish}) even though its quantum correspondent $\wPhi$ 
does not, because of specific features of the density operator $\wrho$. In this 
case, the couples $(\wT,\wspt)$ and $(\wT',\wspt')$ are to be considered equivalent 
only with regard to a particular density operator, that is for a specific quantum
state.

We will see in the next two sections that the equivalence between couples of tensors 
crucially depends on the symmetry properties of the physical state $\wrho$ 
(either mixed or pure). Particularly, we shall see that if $\wrho$ is the usual 
thermodynamical equilibrium operator, proportional to $\exp[-\widehat H/T + \mu 
\widehat Q/T]$, any quantum tensor $\wPhi$ will result in a mean value $\Phi$ 
fulfilling eq.~(\ref{phieq}) and (\ref{vanish}). This means that {\em all} possible 
quantum microscopic stress-energy and spin tensors will yield the same physics in 
terms of macroscopically observable quantities.

\section{Thermodynamical equilibrium}
\label{thermo}

The familiar thermodynamical equilibrium distribution (in the thermodynamical limit 
$V \to \infty$):
\begin{equation}\label{thermal}
\wrho = \frac{1}{Z} \exp(-\widehat H/T + \mu \widehat Q/T)
\end{equation}
where $\widehat Q$ denotes a conserved charge, and $Z$ is the grand-canonical 
partition function
\begin{equation*}
  Z = \tr [\exp(-\widehat H/T + \mu \widehat Q/T)] 
\end{equation*}
is remarkably symmetric. It is space-time translationally invariant, since both 
$\widehat Q$ and $\widehat H$ commute with translation operators $\widehat{\sf T}(a) =
\exp[\ii a \cdot \widehat P]$. This entails that the mean value of any space-time 
dependent operator $\widehat A(x)$, including stress-energy and spin tensor, are
independent of the space-time position:
\begin{equation}\label{genop}
 \tr [\wrho :\! \widehat A(x+a) \!:] = \tr [\wrho :\! \widehat{\sf T}(a) \widehat A(x) 
 \widehat{\sf T}(a)^{-1}\!:] = \tr [\wrho \, \widehat{\sf T}(a) :\! \widehat A(x) 
 \!:\widehat{\sf T}(a)^{-1}] = \tr [ \widehat{\sf T}(a)^{-1} \wrho \, \widehat{\sf T}(a) 
 :\! \widehat A(x) \!:] = \tr [ \wrho :\! \widehat A(x) \!:]
\end{equation}
where the ciclicity of the trace and the transparency of the normal ordering with
respect to translations have been used \footnote{Here a comment is in order. The
transparency of the normal ordering with respect to a conjugation transformation, 
that is $:\! {\sf A} F(\Psi) {\sf A}^{-1}\!: = {\sf A} :\! F(\Psi) \!: {\sf A}^{-1}$ 
where ${\sf A}$ is a translation or a Lorentz transformation and $F$ a function
of the fields and its derivatives, is guaranteed for free fields provided that 
the vacuum $\ket{0}$ is an eigenstate of the same transformation, which is always 
the case. For interacting fields, we will assume that the definition of normal 
ordering (for this problem, see e.g. ref.~\cite{deweert}) is such that transparency 
for conjugation holds; anyhow, for the examined case in Sect.~\ref{dirac} we will just 
need transparency for a free field.}. As a consequence, the mean 
value of any space-time derivative vanishes, and so will do the divergences on the
right hand side of eq.~(\ref{transfq}). Therefore, the mean stress-energy tensor 
will be the same regardless of the particular microscopic quantum tensor used. 
For instance, for the Dirac field, the three tensors:
\begin{equation}\label{dirtens}
  \ii {\overline \Psi} \gamma^\mu \partial^\nu \Psi \qquad \qquad 
  \frac{\ii}{2} {\overline \Psi} \gamma^\mu \codev^{\nu} \Psi \qquad \qquad
  \frac{\ii}{4} \left[ {\overline \Psi} \gamma^\mu \codev^\nu \Psi + 
  (\mu \leftrightarrow \nu) \right]  
\end{equation}  
will result in the same mean stress-energy tensor. 

Also, the density operator (\ref{thermal}) manifestly enjoys rotational symmetry,
for $\widehat H$ and $\widehat Q$ commute with rotation operators $\widehat {\sf R}$.
This implies that most components of tensors vanish. To show that, it is sufficient to 
choose suitable rotational operators and repeat the same reasoning as in 
eq.~(\ref{genop}). For instance, choosing the ${\sf R}_2(\pi)$ operator, i.e. the 
rotation of 180 degrees around the $2$ (or $y$) axis, changing the sign of $1$ (or $x$) 
and $3$ (or $z$) components and leaving 2 and 0 unchanged, in the $x=(t,{\bf 0})$
one has:
\begin{eqnarray}\label{genop2}
 && T^{12}(x) = \tr [\wrho :\! \widehat T^{12}(x) \!:] = \tr [ \widehat{\sf R}_2(\pi) 
 \wrho \, \widehat{\sf R}_2(\pi)^{-1} :\! \widehat T^{12}(x) \!:] = 
 \tr [\wrho \, \widehat{\sf R}_2(\pi)^{-1} :\! \widehat T^{12}(x) \!: \widehat{\sf R}_2(\pi)] 
 \nonumber \\
 = && \tr [ \wrho \, {\sf R}_2(\pi)^1_{\mu} {\sf R}_2(\pi)^2_{\nu} :\! 
 \widehat T^{\mu \nu} ({\sf R}_2(\pi)^{-1}(x)) \!:] =  - \tr [ \wrho :\! \widehat T^{12}
 ({\sf R}_2(\pi)^{-1}(x)) \!:] = - \tr [ \wrho :\! \widehat T^{12}(x) \!:] = -T^{12}(x)
\end{eqnarray}
where, in the last equality, we have taken advantage of the homogeneity of all mean
values shown in eq.~(\ref{genop}); thus, $T^{12}((t,{\bf 0}))=0$ and, in view of the
translational invariance $T^{12}(x)=0 \; \forall x$. Similarly, by choosing other
rotation operators, it can be shown that all off-diagonal elements of a tensor 
vanish. The only non-vanishing components are the diagonal ones, which, again owing
to the rotational symmetry (choose ${\sf R}_i(\pi/2)$ and repeat the above reasoning), 
are equal:
\begin{equation*}
   T^{11}(x) = T^{22}(x) = T^{33}(x)
\end{equation*}      
The component $T^{00}(x)$ can also be non-vanishing and its value is unrelated to the
other diagonal ones. Altogether, the mean stress-energy tensor can only have the 
diagonal (symmetric) form:
\begin{equation*}
 T^{\mu \nu} = 
 \left( \begin{array}{cccc} 
                     \rho  & 0 & 0 & 0  \\
                     0 & p & 0 & 0 \\
		     0 & 0 & p & 0 \\
		     0 & 0 & 0 & p 
  \end{array} \right) = (\rho + p) \hat t^\mu \hat t^\nu - p g^{\mu \nu}
\end{equation*}
where $\hat t$ is the unit time vector with components $(1,{\bf 0})$ and $\rho$ 
and $p$ have the physical meaning of proper energy density and pressure. It should 
be stressed that, for a system at full thermodynamical equilibrium described by 
$\wrho$ in eq.~(\ref{thermal}) they would be the same regardless of the particular 
form of the quantum stress-energy tensor, e.g. those in  eq.~(\ref{dirtens}) for 
the free Dirac field.

As far as the superpotential is concerned, it is easy to convince oneself, by using 
suitable rotations, that the only non-vanishing components are:
\begin{equation*}
   \Phi^{1,01}(x) = \Phi^{2,02}(x) = \Phi^{3,03}(x) = - \Phi^{1,10}(x) = 
   - \Phi^{2,20}(x) = - \Phi^{3,30}(x)
\end{equation*}
Hence, one scalar function $B$, independent of $x$, is sufficient to 
determine the spin tensor for a system at full thermodynamical equilibrium:
\begin{equation*}
       \Phi^{\lambda, \mu \nu} = B (g^{\lambda \nu} \hat t^\mu - 
       g^{\lambda \mu} \hat t^\nu  )
\end{equation*}
This tensor has exactly the form for a ``good" superpotential derived in eq.~(\ref{vanish})
fulfilling condition (\ref{fincond}). In conclusion, {\em any} transformation of the 
kind (\ref{transfq}) will yield the same energy, momentum and angular momentum 
density for all inertial frames and so, all quantum stress-energy and spin 
tensors are equivalent as far as the density operator (\ref{thermal}) is concerned.

\section{Thermodynamical equilibrium with angular momentum}
\label{thermang}

The situation is remarkably different for a thermodynamical system having a macroscopic
non-vanishing total angular momentum. In this case, in its rest frame (defined as
the one where the total momentum vanishes) the density operator reads 
\cite{landau,vilenkin}:
\begin{equation}\label{thermj}
\wrho = \frac{1}{Z_\omega} \exp(-\widehat H/T + \omegav \cdot \widehat{\bf J}/T 
+ \mu \widehat Q/T)
\end{equation}
where $\omegav$ has the physical meaning of a constant, fixed angular velocity 
around which the system rigidly rotates. The factor $Z_\omega$ is the {\em rotational} 
grand-canonical partition function:
\begin{equation}\label{rotpf}
  Z_\omega = \tr [ \exp(-\widehat H/T + \omegav \cdot \widehat{\bf J}/T + 
\mu \widehat Q/T)] 
\end{equation}
The density operator (\ref{thermj}) is much less symmetric than that in (\ref{thermal}) 
and this has remarkable and interesting consequences on the allowed transformations  
of stress-energy and spin tensor. The surviving symmetries in (\ref{thermj}) are
time-translations ${\sf T}(t)$ and translations along the $\omegav$ axis ${\sf T}(z)$, 
rotations around the $\omegav$ axis ${\sf R}_{\hat \omegav}(\varphi)$ and reflection 
$\Pi_{\hat\omegav}$ with respect to planes orthogonal to $\omegav$ (if $\widehat H$ 
is parity-invariant). 

The density operator (\ref{thermj}) can be obtained in several fashions: by maximizing
the entropy with the constraint of fixed mean value of angular momentum \cite{balian},
generalizing to the quantum-relativistic case an argument used by Landau for classical
systems \cite{landau} or as the limiting macroscopic case of a quantum statistical 
system with finite volume and fixed angular momentum in its rest frame in an exact 
quantum sense, i.e. belonging to a specific representation of the rotation group 
\cite{becaferr2}. It should be pointed out that $\widehat H, \widehat Q$ and the 
angular momentum operator along the $\omegav$ direction commute with each other, 
so that the exponential in (\ref{thermj}) also factorizes.

The density operator (\ref{thermj}) implies that, in its rest frame, the system is 
rigidly rotating with a velocity field ${\bf v} = \omegav \times {\bf x}$. The 
classical, non-relativistic derivation by Landau \cite{landau} shows this in a very 
simple fashion by assuming that the system is made of macroscopic cells. To show 
the same thing within a quantum formalism, implies a little more effort, which is 
nevertheless quite enlightening. Consider a vector field $\widehat V(x)$ and 
calculate its mean value at a point $x+a$ by using space-time translation operators. 
Evidently:
\begin{eqnarray}\label{trace1}
 && \tr [\wrho :\! \widehat V^\nu (x+a) \!:] = \tr [\wrho \, \widehat{\sf T}(a) :\! 
 \widehat V^\nu (x) \!:\widehat{\sf T}(a)^{-1}] = \tr [ \widehat{\sf T}(a)^{-1} \wrho \, 
 \widehat{\sf T}(a) :\! \widehat V^\nu (x) \!:] \nonumber \\
 && = \frac{1}{Z_\omega} \tr [ \widehat{\sf T}(a)^{-1} \e^{-\widehat H/T + \omegav 
 \cdot \widehat{\bf J}/T + \mu \widehat Q/T} \widehat{\sf T}(a) :\! \widehat V^\nu(x) \!:] 
 = \frac{1}{Z_\omega} \tr [ \e^{-\widehat H/T + \omegav \cdot \widehat{\sf T}(a)^{-1} 
 \widehat{\bf J} \widehat{\sf T}(a)/T + \mu \widehat Q/T}  :\! \widehat V^\nu (x) \!:]
\end{eqnarray}
where known commutation properties $[\widehat Q, \widehat P^\mu] = 0$ and $[\widehat H, 
\widehat P^\mu] = 0$ have been used. Now, from the theory of Poincar\'e group is known 
that:
\begin{equation}
  \widehat{\sf T}(a)^{-1} \widehat{\bf J} \widehat{\sf T}(a) = \widehat{\bf J}
  + {\bf a} \times \widehat{\bf P}
\end{equation}
whence, from (\ref{trace1}):
\begin{equation}\label{trace2}
 \tr [\wrho :\! \widehat V^\nu (x+a) \!:] = \frac{1}{Z_\omega} \tr [ \e^{-\widehat H/T 
 + \omegav \cdot (\widehat{\bf J} + {\bf a} \times \widehat{\bf P})/T + \mu \widehat Q/T}  
 :\! \widehat V^\nu (x) \!:] =  \frac{1}{Z_\omega} \tr [ \e^{-\widehat H/T 
 + (\omegav \times {\bf a}) \cdot \widehat{\bf P}/T + \omegav \cdot \widehat{\bf J}/T + 
 \mu \widehat Q/T} :\! \widehat V^\nu (x) \!:] 
\end{equation}
We can now define the temperature four-vector: 
\begin{equation*}
\beta = \frac{1}{T} (1, \omegav \times {\bf a})
\end{equation*}
which can also be rewritten as:
\begin{equation}\label{fourv}
  \beta = \frac{1}{T_0} u = \frac{1}{T_0} (\gamma, \gamma {\bf v})  
\end{equation}
where ${\bf v} = \omegav \times {\bf a}$, $\gamma = 1/\sqrt{1 - v^2}$ and $T_0 = 
\gamma T$. The vector ${\bf v}$ is manifestly a rigid velocity field, while $T_0$
is the inverse modulus of $\beta$, i.e. the comoving temperature which differs by
the constant uniform $T$ by a $\gamma$ factor \cite{israel,becatinti}. The mean 
value of $V^\nu (x+a)$ in eq.~(\ref{trace2}) becomes:
\begin{equation}\label{trace3}
 \tr [\wrho :\! \widehat V^\nu (x+a) \!:] = \frac{1}{Z_\omega} \tr [\exp (
 - \beta(a) \cdot \widehat P + \omegav \cdot \widehat{\bf J}/T + \mu 
 \widehat Q/T ) :\! \widehat V^\nu (x) \!: ] 
\end{equation}
Since $\beta$ is timelike (provided that $ v < 1$), it is possible to find a Lorentz
transformation ${\sf \Lambda}$ such that:
\begin{equation}\label{lorentz}
   \frac{1}{T_0} \beta_\mu = u_\mu = {\sf \Lambda}_{0\mu} = g_{0 \lambda} 
   {\sf \Lambda}^\lambda_\mu
\end{equation}
A convenient choice is the pure Lorentz boost along the ${\bf v}= \omegav \times 
{\bf a}$ direction,  which, being ortogonal to $\omegav$, leaves the operator 
$\widehat{\bf J}\cdot \omegav$ invariant:
\begin{equation*}
  {\sf \Lambda} = \exp[ - \ii \, {\rm arccosh}(\gamma) \hat{\bf v} \cdot 
  {\bf {\sf K}}]
\end{equation*}
where ${\sf K}_i \;\; (i=1,2,3)$ are the generators of pure Lorentz boosts. 
Thereby, the trace on the right hand side of the eq.~(\ref{trace3}) can 
be written: 
\begin{eqnarray}\label{trace4}
 && \tr \left[ \e^{-{\sf \Lambda}_{0\mu} \widehat P^\mu + 
 \omegav \cdot \widehat{\bf J}/T + \mu \widehat Q/T} :\! 
 \widehat V^\nu (x) \!: \right] =
  \tr \left[ \e^{- \widehat{\sf \Lambda}^{-1} (\widehat P^0/T_0 
 + \gamma \omegav \cdot \widehat{\bf J}/T_0 + \gamma \mu \widehat Q/T_0) 
 \widehat{\sf \Lambda}} :\! \widehat V^\nu (x) \!: \right] \nonumber \\
 &=& \tr \left[ \widehat{\sf \Lambda}^{-1} \e^{- \widehat P^0/T_0 + \gamma \omegav \cdot 
 \widehat{\bf J}/T_0 + \gamma \mu \widehat Q/T_0} \widehat{\sf \Lambda}:\! 
 \widehat V^\nu (x) \!: \right] =
  \tr \left[ \e^{- \widehat P^0/T_0 + \gamma \omegav \cdot 
 \widehat{\bf J}/T_0 + \gamma \mu \widehat Q/T_0} \widehat{\sf \Lambda}:\! 
 \widehat V(x) \!: \widehat{\sf \Lambda}^{-1} \right] \nonumber \\
 &=& ({\sf \Lambda}^{-1})^\nu_\mu \tr \left[ \e^{- \widehat P^0/T_0 + \gamma \omegav \cdot 
 \widehat{\bf J}/T_0 + \gamma \mu \widehat Q/T_0} :\! 
 \widehat V^\mu({\sf \Lambda}(x)) \!: \right]
\end{eqnarray}
Finally, from (\ref{trace3}) and (\ref{trace4}) we get:
\begin{equation}\label{general}
 \tr [\wrho :\! \widehat V^\nu(x+a) \!:] = \frac{1}{Z_\omega} 
 ({\sf \Lambda}^{-1})^\nu_\mu \tr \left[ \e^{- \widehat P^0/T_0 + \gamma \omegav 
 \cdot \widehat{\bf J}/T_0 + \gamma \mu \widehat Q/T_0} :\! 
 \widehat V^\mu({\sf \Lambda}(x)) \!: \right]
\end{equation}
which tells us how to calculate the mean value of a vector field at any space-time
point given its value in some other specific point. The most interesting feature of
eq.~(\ref{general}) is that the density operator on the right hand side the same as
$\wrho$ on the left hand side with the replacement:
\begin{equation}\label{repla}
 T \to T_0 = \gamma(a) T  \qquad  \omegav \to \gamma(a) \omegav  
 \qquad \mu \to \gamma(a)\mu
\end{equation}

\begin{center}
\begin{figure}[ht]
\epsfxsize=2.5in
\epsffile{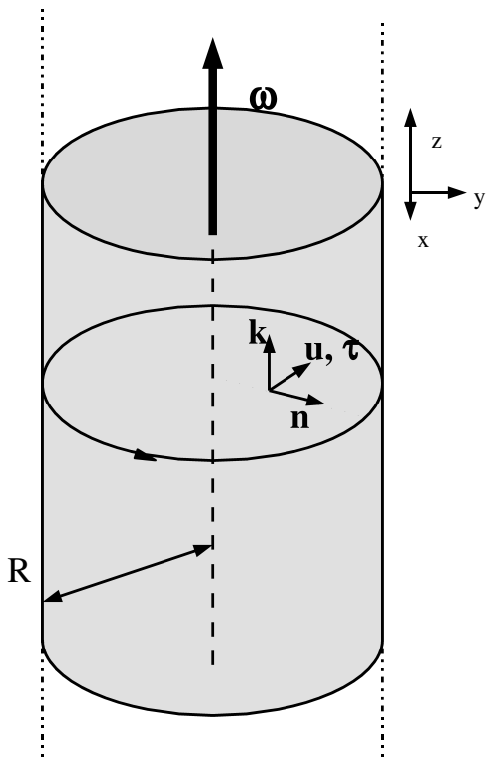} 
\caption{Rotating cylinder with finite radius $R$ at temperature $T$. Also shown
the inertial frame axes and the spatial parts of the vectors of tetrad (\ref{tetrad}).
\label{illu}}
\end{figure}
\end{center}

If we choose $x =(0,{\bf 0})$, i.e. the origin of Minkowski coordinates, and
$a = (0,{\bf a})$ the eq.~(\ref{general}) implies:
\begin{equation}\label{vboost}
 \tr [\wrho :\! \widehat V^\nu(0,{\bf a}) \!:] = \frac{1}{Z_\omega} 
 ({\sf \Lambda}^{-1})^\nu_\mu \tr \left[ \e^{- \widehat P^0/T_0 + \gamma \omegav 
 \cdot \widehat{\bf J}/T_0 + \gamma \mu \widehat Q/T_0} :\! 
 \widehat V^\mu(0,{\bf 0}) \!: \right]
\end{equation}
that is the mean value of the vector field at any space-time point (it should be 
kept in mind that $\wrho$ is invariant by time translation and so any mean value 
is stationary) is completely determined by the mean value at the origin of the 
coordinates, with the same density operator, modulo the replacement of thermodynamical 
parameters in (\ref{repla}). This particular value is strongly constrained by the 
symmetries of $\wrho$. Let us identify the $\omegav$ direction as that of the $z$ 
(or $3$) axis (see fig.~\ref{illu}) and consider the reflection ${\sf \Pi}_z$ with 
respect to $z=0$ plane and the rotation ${\sf R}_3(\pi)$ of an angle $\pi$ around 
the $z$ axis; by repeating 
the same reasoning as for eq.~(\ref{genop2}) for $V^\nu (0)$ we can easily conclude 
that the time component $V^0(0)$ is the only one having a non-vanishing mean value. 
Note, though, that the mean value on the right-hand side of (\ref{vboost}) depends 
on the distance $r$ from the axis because the density operator is modified by the
replacement of the uniform temperature $T$ with a radius-dependent $T_0 = \gamma T$. 
Therefore, according to eq.~(\ref{vboost}) and using (\ref{lorentz}), the mean value 
of the vector field can be written:
\begin{equation}\label{vector}
 V_\nu(x) = \tr [\wrho :\! \widehat V_\nu (x) \!:] =
\frac{1}{Z_\omega} ({\sf \Lambda}^{-1})_{\nu 0} 
 \tr \left[ \e^{- \widehat P^0/T_0(r) + \gamma(r)\omegav \cdot \widehat{\bf J}/T_0(r) 
 + \gamma(r)\mu\widehat Q/T_0(r)} :\! \widehat V^0(0) \!: \right]
 \equiv {\sf \Lambda}_{0 \nu} V(r) = V(r) u_\nu   
\end{equation}
i.e. it must be collinear with the four-velocity field $u=(\gamma, \gamma {\bf v})$
in eq.~(\ref{fourv}) and, therefore, its field lines are circles centered on the 
$z$ axis and orthogonal to it. Similarly, we can obtain the general form of tensor 
fields of various rank and specific symmetry properties as a function of the basic 
four-velocity field. 

However, the previous derivation relies on the fact that the system is infinitely
extended in space. Indeed, at a distance from the axis such that $|\omegav \times
{\bf x}|= 1$ the velocity becomes equal to the speed of light and the system has a
singularity. We cannot, therefore, take the strict thermodynamical limit $V \to 
\infty$ for a system with macroscopic angular momentum. Instead, we have to enforce 
a spatial cut-off at some distance and figure out how this reflects on the most
general forms of vector and tensor fields.

Enforcing a bounded region $V$ for a thermodynamical system implies the replacement 
of all traces over the full set of states with a trace over a complete set of states 
$\ket{h_V}$ of the fields for this region $V$, that we indicate with $\tr_V$:
$$
  \tr \to \tr_V  = \sum_{h_V} \bra{h_V} \ldots \ket{h_V}
$$
The density operator is the same as in (\ref{thermj}) with the partition function
now obtained by tracing over the localized states. It may sometimes be convenient 
to introduce the projection operator:
$$
  \Pro_V = \sum_{h_V} \ket{h_V}  \bra{h_V}
$$
which allows to maintain the trace over the full set of states of provided that we
replace $\wrho$ with $\Pro_V \wrho $; for a generic operator $\widehat A$
\begin{equation*}
  \tr_V [\wrho \, \widehat A] = \tr [\Pro_V \wrho \, \widehat A]
\end{equation*}
which amounts to state that the effective density operator is now $\wrho_V$:
\begin{equation}\label{thermjv}
  \wrho_V = \frac{1}{Z_\omega} \Pro_V 
  \exp(-\widehat H/T + \omegav \cdot \widehat{\bf J}/T + \mu \widehat Q/T) 
\end{equation}
where:
\begin{equation*}  
 Z_\omega = \tr [ \Pro_V \exp(-\widehat H/T + \omegav \cdot \widehat{\bf J}/T + 
 \mu \widehat Q/T) ] = \tr_V [ \exp(-\widehat H/T + \omegav \cdot \widehat{\bf J}/T 
 + \widehat Q/T) ]
\end{equation*}
In order to maintain the same symmetry of the density operator in (\ref{thermj}), 
$\Pro_V$ has to commute with $\widehat J_z$, $\widehat H$, $\widehat P_z$, the
Lorentz boost along $z$ $\widehat K_z$ and the reflection operator with respect 
to any plane parallel to $z=0$, $\widehat{\sf \Pi}_z$ (see fig.~\ref{illu}). These 
requirements are met if the region $V$ is a static longitudinally indefinite 
cylinder with finite radius $R$ and axis $\omegav$, and we will henceforth take 
this assumption. 

There are two important consequences of having a finite radius $R$. Because of 
the presence of the projector $\Pro_V$, the previous derivation which led us to 
express the vector field according to the simple formula (\ref{vector}) cannot 
be carried over to the case of finite (though macroscopic) radius. The reason
is that $\Pro_V$ does not commute with the Lorentz boost along ${\bf v}$ or,
otherwise stated, a Lorentz boost along a direction other than $z$ will not 
transform the set of states $\ket{h_V}$ into themselves, as needed for completeness.
So, one the one of the crucial steps in eq.~(\ref{trace4}) no longer holds and 
specifically:
\begin{equation*}  
 \tr \left[ \Pro_V \widehat{\sf \Lambda}^{-1} \e^{- \widehat P^0/T_0 + 
 \gamma \omegav \cdot \widehat{\bf J}/T_0 + \gamma \mu \widehat Q/T_0} 
 \widehat{\sf \Lambda}:\! \widehat V^\nu (x) \!: \right] \ne 
  \tr \left[ \Pro_V \e^{- \widehat P^0/T_0 + 
 \gamma \omegav \cdot \widehat{\bf J}/T_0 + \gamma \mu \widehat Q/T_0} 
 \widehat{\sf \Lambda}:\! \widehat V^\nu (x) \!: \widehat{\sf \Lambda}^{-1}\right]
\end{equation*}
As a consequence, general vector and tensor fields will be more complicated than 
in the unphysical infinite radius case and get additional components. The most
general expressions of mean value of fields in the cases of interest for the 
stress-energy and spin tensor will be systematically determined in the next section.
The second consequence is that boundary conditions for the quantum fields must
be specified at a finite radius value $R$, but we will see that those conditions
alone cannot ensure the validity of eqs.~(\ref{phieq}) or (\ref{vanish}), which
are local conditions.

\section{Tensor fields in an axisymmetric system}
\label{axisymm}

In this section we will write down the most general forms of vector and tensor fields
in an axisymmetric system, i.e. a system with the same symmetry features of the 
thermodynamical rotating system at equilibrium studied in the previous section. 
The goal of this section is to establish the conditions, if any, to be fulfilled 
by the superpotential to generate a good transformation of the stress-energy
and spin tensors. 

\subsection{Vector field}

The decomposition of a vector field will serve as a paradigm for more complicated 
cases. The idea is to take a suitable tetrad of space-time dependent orthonormal 
four-vectors and decompose the vector field onto this basis. The tetrad we choose is 
dictated by the cylindrical symmetry:
\begin{equation}\label{tetrad}
  u = (\gamma, \gamma {\bf v}) \qquad \tau = (\gamma v, \gamma\hat{\bf v}) \qquad
  n = (0, \hat{\bf r}) \qquad k = (0, \hat{\bf k})
\end{equation}
where $\hat {\bf r}$ is the radial versor in cylindrical coordinates, while 
$\hat{\bf k}$ is the versor of the $z$ axis, that is the axis of the cylinder
(see fig.~\ref{illu}). 

Due to symmetry for reflections with respect to $z=const$ planes, the most general 
vector field $V$ has vanishing component on $k$, and therefore:
\begin{equation}\label{vectorfield}
  V = A(r) u + B(r) \tau + C(r) n
\end{equation}
where $A, B, C$ are scalar functions which can only depend on the radial coordinate
$r$, owing to the cylindrical symmetry. Note the presence of two additional components  
with respect to the infinitely extended cylinder case in eq.~(\ref{vector}).
For symmetry reasons the only surviving component of the field at the axis is the 
time component, so $B(0)=C(0)=0$.

If the field is divergence-free, then $C(r) \equiv 0$.

\subsection{Rank 2 antisymmetric tensor field}

Any antisymmetric tensor field of rank 2 can be decomposed first as:
\begin{equation*}
 A^{\mu \nu} = \epsilon^{\mu \nu \rho \sigma} X_\rho u_\sigma + 
 Y^\mu u^\nu - Y^\nu u^\mu 
\end{equation*}
where:
\begin{equation*}
 X^\rho = -\frac{1}{2} \epsilon^{\rho \alpha \beta \gamma} A_{\alpha \beta} u_\gamma
 \qquad Y^\rho = A^{\rho \alpha} u_\alpha
\end{equation*}
and, thus, $X$ and $Y$ are two space-like vector fields such that $X \cdot u = Y 
\cdot u =0$. Because of the reflection symmetry with respect to $z=const$ planes, one 
has $A_{xz}=A_{yz}=A_{0z}=0$ and this in turn entails that, being $u_z = 0$, the
only non-vanishing component of the pseudo-vector $X$ is along $k$. Conversely, $Y$ 
is a polar vector and it has components along $\tau$ and $n$ which must vanish 
in $r=0$. Altogether:
\begin{equation}\label{anti}
 A^{\mu \nu} = D(r) \epsilon^{\mu \nu \rho \sigma} k_\rho u_\sigma + 
 E(r) (\tau^\mu u^\nu - \tau^\nu u^\mu) + F(r) (n^\mu u^\nu - n^\nu u^\mu) 
\end{equation}
with $E(0)=F(0)=0$. Since:
\begin{equation*}
  \epsilon^{\mu \nu \rho \sigma} k_\rho u_\sigma = n^\mu \tau^\nu - n^\nu \tau^\mu 
\end{equation*}
(which can be easily checked), the expression (\ref{anti}) can be rewritten as:
\begin{equation}\label{anti3}
 A^{\mu \nu} = D(r) (n^\mu \tau^\nu - n^\nu \tau^\mu)  + 
 E(r) (\tau^\mu u^\nu - \tau^\nu u^\mu) + F(r) (n^\mu u^\nu - n^\nu u^\mu) 
\end{equation}
%

\subsection{Rank 2 symmetric tensor field}

For the symmetric tensor $S^{\mu \nu}$ we will employ an iteration method in order
to write down the most general decomposition. First, we project the tensor onto the
$u$ field:
\begin{equation*}
  S^{\mu \nu} = G(r) u^\mu u^\nu + q^\mu u^\nu + q^\nu u^\mu + \Theta^{\mu \nu}
\end{equation*}
where $q \cdot u = 0$ and $\Theta^{\mu \nu} u_\nu = 0$. Then, we decompose the 
space-like polar vector field $q$ according to (\ref{vectorfield})
\begin{equation*}
   q = H(r) \tau + I(r) n
\end{equation*}
with $H(0)=I(0)=0$, and we project the tensor $\Theta$ in turn onto the vector 
field $\tau$:
\begin{equation*}
  S^{\mu \nu} = G(r) u^\mu u^\nu + H(r) (\tau^\mu u^\nu + \tau^\nu u^\mu) +
  I(r) (n^\mu u^\nu + n^\nu u^\mu) + J(r) \tau^\mu \tau^\nu + h^\mu \tau^\nu
  + h^\nu \tau^\nu + \Xi^{\mu \nu}
\end{equation*}
being $h \cdot u = h \cdot \tau = 0$ (whence $h = K(r) n$ with $K(0)=0$) and 
$\Xi^{\mu \nu} \tau_\nu = \Xi^{\mu \nu} u_\nu = 0$. This procedure can be 
iterated projecting $\Xi$ onto $n$ and the thus-obtained new symmetric tensor 
onto $k$. Thereby, we get:
\begin{eqnarray}\label{symm1}
 && S^{\mu \nu} = G(r) u^\mu u^\nu + H(r) (\tau^\mu u^\nu + \tau^\nu u^\mu) +
  I(r) (n^\mu u^\nu + n^\nu u^\mu) \nonumber \\
 && + J(r) \tau^\mu \tau^\nu + K(r) (n^\mu 
  \tau^\nu + n^\nu \tau^\mu) + L(r) n^\mu n^\nu + M(r) k^\mu k^\nu 
\end{eqnarray}  
However, since $u^\mu u^\nu - \tau^\mu \tau^\nu - n^\mu n^\nu - k^\mu k^\nu
= g^{\mu \nu}$ the last term can be replaced with a linear combination of
all other diagonal terms plus a term in $g^{\mu \nu}$ and the most general
symmetric tensor can be rewritten, after a suitable redefinition of the 
scalar coefficients, as:
\begin{eqnarray*}
  && S^{\mu \nu} = G(r) u^\mu u^\nu + H(r) (\tau^\mu u^\nu + \tau^\nu u^\mu) +
  I(r) (n^\mu u^\nu + n^\nu u^\mu) \nonumber \\
  && + J(r) \tau^\mu \tau^\nu + K(r) (n^\mu 
  \tau^\nu + n^\nu \tau^\mu) + L(r) n^\mu n^\nu - M(r) g^{\mu \nu} 
\end{eqnarray*}  
where $H(0)=I(0)=K(0)=0$.

\subsection{Rank 3 spin-like tensor field}

The decomposition of a rank 3 tensor is carried out in an iterative way, similarly 
to what we have just done for the rank 2 symmetric tensor. First, we project the 
tensor onto the vector $u$ and, taking the antisymmetry of $\mu \nu$ indices 
into account, one obtains:
\begin{equation}\label{phi1}
 \Phi^{\lambda, \mu \nu} = u^\lambda (f^\mu u^\nu - f^\nu u^\mu) +
 u^\lambda \Gamma^{\mu \nu} + \Sigma^{\lambda \mu} u^\nu - \Sigma^{\lambda \nu}
 u^\mu + \Upsilon^{\lambda, \mu \nu}
\end{equation}
where all vector and tensor fields have vanishing contractions with $u$ for
any index. Particularly, using the general expressions (\ref{vectorfield}) 
and (\ref{anti3}), the vector field $f$ and the antisymmetric tensor $\Gamma$ 
read:
\begin{equation}\label{gamma}
  f = E(r) \tau^\mu + F(r) n^\mu \qquad \qquad \Gamma^{\mu \nu} 
  = D(r) (n^\mu \tau^\nu - n^\nu \tau^\mu)  
\end{equation}  
with $E(0)=F(0)=0$. The tensor $\Sigma$ can be decomposed as the sum of a symmetric 
and an antisymmetric part; having vanishing contractions with $u$, according
to eqs.~(\ref{anti3}) and (\ref{symm1}), it can be written as:
\begin{equation}\label{sigma}
 \Sigma^{\lambda \mu} = N(r) (n^\lambda \tau^\mu - n^\mu \tau^\lambda)
 + P(r) \tau^\lambda \tau^\mu + Q(r) (n^\lambda \tau^\mu + n^\mu \tau^\lambda) + 
   R(r) n^\lambda n^\mu + S(r) k^\lambda k^\mu
\end{equation}
wiht $Q(0)=0$. 
The tensor $\Upsilon$ is projected in turn onto $n$ and the above procedure is 
iterated. Then, similarly to eq.~(\ref{phi1}):
\begin{equation}\label{upsilon}
 \Upsilon^{\lambda, \mu \nu} = \chi^{\mu \nu} n^\lambda + (h^\mu n^\nu - h^\nu n^\mu)
 n^\lambda + \Theta^{\lambda \mu} n^\nu - \Theta^{\lambda \nu} n^\mu + 
 \Lambda^{\lambda, \mu \nu}
\end{equation}
where all tensors have vanishing contractions with $u$ and $n$. 
The antisymmetric tensor $\chi$ must be orthogonal to $u$ and $n$ and, therefore, 
according to eq.~(\ref{anti3}), vanishes. On the other hand, the vector field $h$ 
can only have non-vanishing component on $\tau$ and so $h = T(r) \tau$ with $T(0)=0$. 
Finally, the tensor $\Theta$ must be orthogonal to $n$, besides $u$, hence, using 
eqs.~(\ref{anti3}) and(\ref{symm1}), can only be of the form:
\begin{equation}\label{theta}
 \Theta^{\lambda \mu} = U(r) \tau^\lambda \tau^\mu + V(r) k^\lambda k^\mu
\end{equation}
Likewise, the tensor $\Lambda$ can be decomposed onto $\tau$ and, because of vanishing 
contractions with $u$ and $n$, it can be written as:
\begin{equation}\label{lambda}
 \Lambda^{\lambda, \mu \nu} = W(r) k^\lambda (k^\mu \tau^\nu - k^\nu \tau^\mu) 
\end{equation}

Putting together eqs.~(\ref{phi1}), (\ref{gamma}), (\ref{sigma}), (\ref{upsilon}),
(\ref{theta}) and (\ref{lambda}), the general decomposition of a rank 3 tensor with 
antisymmetric $\mu\nu$ indices is obtained:
\begin{eqnarray}\label{phifin}
 && \Phi^{\lambda, \mu \nu} = D(r) (n^\mu \tau^\nu - n^\nu \tau^\mu) u^\lambda
 + E(r) (\tau^\mu u^\nu - \tau^\nu u^\mu) u^\lambda + F(r)(n^\mu u^\nu - n^\nu 
 u^\mu) u^\lambda + N(r)(n^\lambda \tau^\mu - n^\mu \tau^\lambda) u^\nu \nonumber \\
 && - N(r) (n^\lambda \tau^\nu - n^\nu \tau^\lambda) u^\mu
 + P(r) \tau^\lambda (\tau^\mu u^\nu - \tau^\nu u^\mu) 
 + Q(r) (n^\lambda \tau^\mu + n^\mu \tau^\lambda) u^\nu
 - Q(r) (n^\lambda \tau^\nu + n^\nu \tau^\lambda) u^\mu \nonumber \\
 && + R(r) n^\lambda (n^\mu u^\nu - n^\nu u^\mu) + S(r) k^\lambda (k^\mu u^\nu
 - k^\nu u^\mu) + T(r) (\tau^\mu n^\nu - \tau^\nu n^\mu) n^\lambda \nonumber \\
 && + U(r) \tau^\lambda (\tau^\mu n^\nu - \tau^\nu n^\mu)
 + V(r) k^\lambda (k^\mu n^\nu - k^\nu n^\mu) + W(r) k^\lambda (k^\mu \tau^\nu 
 - k^\nu \tau^\mu) 
\end{eqnarray}
with $E(0)=F(0)=Q(0)=T(0)=0$.

\begin{center}
\line(1,0){100}
\end{center}

We are now in a position to find out the conditions to be fulfilled by the 
superpotential $\Phi$ to be a good transformation of the stress-energy and spin 
tensors in a thermodynamically equilibrated system with angular momentum, as derived 
at the end of Sect.~\ref{transform}.

Let us start from eq.~(\ref{vanish}), which is the most constraining. Since:
\begin{equation*}
  u^\lambda u^\mu - \tau^\lambda \tau^\mu - n^\lambda n^\mu - k^\lambda k^\mu
= g^{\lambda \mu}
\end{equation*}
we can write a rank 3 tensor (\ref{phifin}) in the form of eq.~(\ref{vanish}) as
long as:
\begin{eqnarray}\label{condition1}
  && V(r) = U(r) = F(r) \nonumber \\
  && P(r) = R(r) = S(r) \nonumber \\
  && E(r) = W(r) = -T(r) \nonumber \\
  && D(r) = N(r) = Q(r) = 0
\end{eqnarray}
which are definitely non-trivial conditions. If these are fulfilled, then the 
superpotential (\ref{phifin}) reduces to:
\begin{equation*}
 \Phi^{\lambda,\mu\nu} = (F(r)n^\mu+E(r)\tau^\mu+P(r)u^\mu) g^{\lambda\nu}
 -(F(r)n^\nu+E(r)\tau^\nu+P(r)u^\nu) g^{\lambda\mu} \equiv K^\mu g^{\lambda\nu}
 - K^\nu g^{\lambda\mu}
\end{equation*}
Now, the field $K^\mu = F(r)n^\mu+E(r)\tau^\mu+P(r)u^\mu$ ought to be a constant 
one, according to eq.~(\ref{fincond}). Since its divergence vanishes, then $F(r)=0$
and, by using the definitions (\ref{tetrad}), we readily obtain the conditions:
\begin{equation}\label{condition2}
  F(r) = 0 \qquad  P(r)/\gamma = {\rm const} \qquad  E(r) = -P(r) \omega r
\end{equation}  
In conclusion, only if a quantum superpotential is such that its mean value, calculated
with the density operator (\ref{thermjv}), fulfills conditions (\ref{condition1}) 
and (\ref{condition2}), is the corresponding transformation (\ref{transfq}) possible.
Otherwise, the original and transformed stress-energy and spin tensors are inequivalent 
because they imply different values of mean energy, momentum or angular momentum 
densities. Since the most general form of the mean value of the superpotential, i.e. 
eq.~(\ref{phifin}) is highly non-trivial, the inequivalence will occur far more often 
than equivalence. To demonstrate this, we will consider a specific instance involving 
the most familiar quantum field endowed with a spin tensor.

\section{An example: the free Dirac field}
\label{dirac}

We now come to the possibly most significant result obtained in this work: the proof
of a concrete instance of inequivalence, involving the simplest quantum field theory 
endowed with a spin tensor, namely the free Dirac field. 

It is well known that the from the lagrangian density: 
\begin{equation}\label{lagra}
  {\cal L}= \frac{\ii}{2} \Psibar \gamma^\mu \codev_\mu \Psi - m \Psibar \Psi
\end{equation}
one obtains, by means of the Noether theorem, the canonical stress-energy and spin 
tensors \cite{zuber}:
\begin{eqnarray}\label{canonical}
 && \wT^{\mu \nu} = \frac{\ii}{2} \Psibar \gamma^\mu \codev^\nu \Psi  \nonumber \\
 && \wspt^{\lambda, \mu \nu} = \frac{1}{2} \Psibar \{ \gamma^\lambda, 
 \Sigma^{\mu \nu} \} \Psi = \frac{i}{8}\Psibar \{\gamma^\lambda,
 [\gamma^\mu,\gamma^\nu] \} \Psi 
\end{eqnarray}
where:
\begin{equation*}
    \Sigma_{ij} =  \epsilon_{ijk} \left( \begin{array}{cc} 
                     \sigma_k/2 & 0  \\
		     0 & \sigma_k/2  \\ 
                                         \end{array} \right)
\end{equation*}    
and $\sigma_k$ are Pauli matrices. The spin tensor obeys the equation (\ref{conservq}):
\begin{equation*}
   \partial_\lambda \wspt^{\lambda, \mu \nu} = \wT^{\nu \mu} - \wT^{\mu \nu}
   =  \frac{\ii}{2} \Psibar \gamma^\nu \codev^\mu \Psi - \frac{\ii}{2} \Psibar \gamma^\mu 
   \codev^\nu \Psi 
\end{equation*}

The couple of quantum tensors in (\ref{canonical}) can be changed through the 
psuedo-gauge transformation in eq.~(\ref{transfq}). Accordingly, if we take $\wPhi=\wspt$, 
namely the superpotential as the original spin tensor itself, a symmetrized 
stress-energy tensor and a vanishing spin tensor are obtained: 
\begin{eqnarray}\label{belinfante}
 && \wT'^{\mu \nu} = \frac{i}{4} \left[ \Psibar \gamma^\mu \codev^\nu \Psi +  
 \Psibar \gamma^\nu \codev^\mu \Psi \right] \nonumber \\
 && \wspt'^{\lambda, \mu \nu} = 0
\end{eqnarray}
This transformation is well known as Belinfante's symmetrization procedure.

One may wonder whether these tensors, fulfilling continuity equations, still exist 
in a bounded region breaking the global translational and Lorentz symmetry, such 
as our cylinder with finite radius; or if, because of the boundary, they get 
additional terms with respect to the usual form. The problem of Dirac field with 
boundary has been tackled and solved by the authors of the MIT bag model \cite{bag}. 
First of all, it should be pointed out that the continuity equations (\ref{conservq}) 
certainly apply to tensors (\ref{canonical}) and (\ref{belinfante}) on-shell, i.e. 
for fields obeying the free Dirac equation within the cylinder. Furthermore, it is 
possible to find suitable boundary conditions, discussed in the next subsection, such 
that the fluxes (\ref{flux}) vanish, as needed, without introducing an ad-hoc discontinuity
in the Dirac field. Thereby, the stress-energy and spin tensors retain the same
form as in the usual no-boundary case and the integrals over the bounded region of 
the time components have the same physical meaning of conserved generators.
It is also possible to derive field equations and canonical tensors (\ref{canonical})
from an action, what is shown in Appendix A.

The spin tensor in eq.~(\ref{canonical}) has a remarkable feature which makes it
easier to check the equivalence of the two couples in eq.~(\ref{canonical}) and 
(\ref{belinfante}): because of the special properties of gamma matrices, the 
spin tensor is also antisymmetric in the first two indices: 
\begin{equation}\label{extranti}
   \wspt^{\lambda, \mu \nu} = - \wspt^{\mu, \lambda \nu}
\end{equation}
and thus the mean value of this tensor is greatly simplified. The antisymmetry in 
the indices $(\lambda,\mu)$ dictates that all coefficients of symmetric $\lambda\mu$ 
terms of the general form of this kind of tensor found in eq.~(\ref{phifin}) vanish:
\begin{equation*}
  E(r)=F(r)=P(r)=Q(r)=R(r)=S(r)=T(r)=U(r)=V(r)=W(r)=0
\end{equation*}
and that $D(r)=N(r)$, so that $\spt$ is simply given by:
\begin{equation}\label{sptpara0}
  \spt^{\lambda,\mu \nu} = D(r) [(n^\mu \tau^\nu - n^\nu \tau^\mu) u^\lambda
  + (n^\lambda \tau^\mu - n^\mu \tau^\lambda) u^\nu - (n^\lambda \tau^\nu - 
 n^\nu \tau^\lambda) u^\mu ]
\end{equation}
and it is described by just one unknown radial function $D(r)$. Therefore, according 
to the conditions (\ref{condition1}), the Belinfante tensors (\ref{belinfante}) 
are equivalent to the canonical ones (\ref{canonical}) only if $D(r)=0$, i.e. 
only if the spin tensor has a vanishing mean value.

For $\lambda=0$, eq.~(\ref{sptpara0}) reads:
\begin{equation*}
 \spt^{0,\mu \nu} = D(r) [(n^\mu \tau^\nu - n^\nu \tau^\mu) u^0
  - \tau^0 (n^\mu u^\nu - n^\nu u^\mu)]  
\end{equation*}
and, because of the antisymmetry, the only non-vanishing components are those with 
both $\mu$ and $\nu$ equal to 1,2,3, indices that we denote with $i,j$. We can then 
write, using (\ref{tetrad}):
\begin{eqnarray}\label{sptpara}
 \spt^{0,ij} &=& D(r) [(n^i \tau^j - n^j \tau^i) u^0 - \tau^0 (n^i u^j - n^j u^i)]
 = D(r) [\gamma^2 (n^i \hat v^j - n^j \hat v^i) - \gamma^2 v^2 (n^i \hat v^j - 
 n^j \hat v^i) \nonumber \\
 &=& D(r) (n^i \hat v^j - n^j \hat v^i) = D(r) \epsilon_{ijk} \hat k^k
\end{eqnarray}
Therefore, as expected, the time part of the spin tensor, contributing to the 
angular momentum density, is equivalent to a pseudo-vector field ${\bf D}(r)$ 
directed along $z$ axis.

According to eq.~(\ref{transf}), the variation of energy-momentum density reads:
\begin{equation}\label{deltamom}
  \frac{1}{2} \partial_\alpha \left( \Phi^{\alpha,0 \nu} - \Phi^{0,\alpha \nu} - 
 \Phi^{\nu, \alpha 0}\right) = \frac{1}{2} \partial_\alpha \left( \spt^{\alpha,0 \nu} 
 - \spt^{0,\alpha \nu} - \spt^{\nu, \alpha 0}\right) = - \frac{1}{2} 
  \partial_\alpha \spt^{0,\alpha \nu}
\end{equation}
which implies at once that the energy density is unchanged because $\spt^{0,\alpha 0} 
=0$ in view of (\ref{extranti}), whereas the momentum density varies by a derivative. 
Using (\ref{sptpara}) and recalling the expression of curl in cylindrical coordinates:
\begin{eqnarray}\label{momdens} 
  T^{0i}_{\rm Belinfante} &=& T^{0i}_{\rm canonical} - \frac{1}{2} \partial_\alpha 
  \spt^{0,\alpha i} = T^{0i}_{\rm canonical} - 
  \frac{1}{2} \partial_\alpha \epsilon_{\alpha ik} D(r) \hat k^k \nonumber \\
  &=& T^{0i}_{\rm canonical} + \frac{1}{2} ({\rm rot}{\bf D})^i = 
  T^{0i}_{\rm canonical} -\frac{1}{2} \frac{\di D(r)}{\di r} {\bf \hat{v}}^i
\end{eqnarray}
Note that this last equation implies that the mean value of the canonical stress-energy
tensor of the Dirac field has a non-trivial antisymmetric part if $D'(r) \ne 0$ 
as, according to (\ref{conservq}): 
\begin{equation*}
 \partial_\alpha \spt^{0,\alpha i} = - \partial_\alpha \spt^{\alpha,0 i}
 = T^{0i}-T^{i0}
\end{equation*}
Now we can write the angular momentum density variation by thermal-averaging 
eq.~(\ref{transfq2}) with $\wPhi=\wspt$:
\begin{eqnarray*}
 {\cal J}^{0, \mu \nu}_{\rm Belinfante} &=& 
 {\cal J}^{0, \mu \nu}_{\rm canonical} 
 + \frac{1}{2} \partial_\alpha \left[ x^\mu \left( \Phi^{\alpha, 0 \nu } - 
 \Phi^{0, \alpha \nu} - \Phi^{\nu, \alpha 0} \right) - x^\nu \left( 
 \Phi^{\alpha, 0 \mu } - \Phi^{0, \alpha \mu} - \Phi^{\mu, \alpha 0} \right) \right]
 \nonumber \\
 &=& \frac{1}{2} \left[ x^\mu \partial_\alpha \left( \Phi^{\alpha, 0 \nu } - 
 \Phi^{0, \alpha \nu} - \Phi^{\nu, \alpha 0} \right) + \Phi^{\mu, 0 \nu } - 
 \Phi^{0, \mu \nu} - \Phi^{\nu, \mu 0} - ( \mu \leftrightarrow \nu) \right]
\end{eqnarray*} 
The sum of all terms linear in the superpotential returns a $-\Phi^{0,\mu\nu}$
(see eq.~(\ref{vanish})) while for the derivative terms we can use eq.~(\ref{deltamom}):
\begin{eqnarray}\label{angmomdens}
 {\cal J}^{0, \mu \nu}_{\rm Belinfante} &=& 
 {\cal J}^{0, \mu \nu}_{\rm canonical} + \frac{1}{2} \left[ x^\mu 
 \partial_\alpha \left( \Phi^{\alpha, 0 \nu } - \Phi^{0, \alpha \nu} - 
 \Phi^{\nu, \alpha 0} \right) - ( \mu \leftrightarrow \nu) \right] - \Phi^{0,\mu \nu}
 \nonumber \\
 &=& {\cal J}^{0, \mu \nu}_{\rm canonical} - \frac{1}{2} \left[ x^\mu 
 \partial_\alpha \spt^{0,\alpha \nu} - x^\nu \partial_\alpha \spt^{0,\alpha \mu} 
 \right] - \spt^{0,\mu \nu}
\end{eqnarray}
Therefore, by plugging the expression of the mean value of the spin tensor in 
eq.~(\ref{sptpara}), the angular momentum pseudo-vector corresponding to the angular 
momentum density in (\ref{angmomdens}) can be finally written:
\begin{equation}\label{angmomdens2}
 {\boldsymbol{\cal J}}_{\rm Belinfante} = {\boldsymbol{\cal J}}_{\rm canonical} 
 - \frac{1}{2} \left( {\bf x} \times \frac{\di D(r)}{\di r} \hat{\bf v} \right)
 - {\bf D}(r) = {\boldsymbol{\cal J}}_{\rm canonical} - \left( \frac{1}{2} r 
 \frac{\di D(r)}{\di r} + D(r) \right) \hat{\bf k}
\end{equation}
In order for the canonical and Belinfante tensors to be equivalent, as has been
mentioned and as it is apparent from eqs.~(\ref{momdens}) and (\ref{angmomdens2}) 
the function $D(r)$ ought to vanish everywhere. If $D'(r) \ne 0$, the two stress-energy 
tensors give two different momentum densities and are thus inequivalent; if, on top of that, 
$D'(r)\ne -2D(r)/r$ then the angular momentum densities are inequivalent as well. 
In the rest of this section we will prove that this is exactly the case, i.e. neither 
of these conditions is fulfilled. In order to show that this is not a problem 
arising from peculiar values of the field at the boundary, we will conservatively 
enforce boundary conditions such that the {\em total} energy, momentum and angular 
momentum operators obtained by integrating the fields within the cylinder are invariant 
by transformation (\ref{transfq}). Note that for this to be true, in the case under
consideration, it is necessary that the function $D(r)$ vanishes at the boundary,
i.e. $D(R)=0$, because the difference between total angular momenta is:
\begin{equation*}
 \int_V \di^3 \x \; \left( 
 {\boldsymbol{\cal J}}_{\rm Belinfante} - {\boldsymbol{\cal J}}_{\rm canonical}\right) 
 = - \int_V \di z \, \di \varphi \, \di r \; r \; 
 \left( \frac{1}{2} r \frac{\di D(r)}{\di r} + D(r) \right) \hat{\bf k} =
 - 2\pi \int_{-\infty}^{+\infty} \!\!\! \di z \int_0^R \di r \; \frac{\di}{\di r} 
 \left( \frac{r^2}{2} D(r) \right) \hat{\bf k}
\end{equation*}

Thereby, we will demonstrate that, although the stress-energy and spin tensors in 
(\ref{canonical}) and (\ref{belinfante}) lead to the same quantum generators, their 
respective mean densities are inconsistent. The problem we are facing is then to 
solve the Dirac equation within a cylinder with finite radius and second-quantize the field.

\subsection{The Dirac field in a cylinder}

The problem of the Dirac field within a cylinder with finite radius has been tackled 
by several authors in the context of the MIT bag model \cite{greiner}. We first
stress that, as we have done thus far, we take the viewpoint of an external inertial 
observer in Minkowski spacetime, seeing the spinning cylinder globally at rest. 
This observer can use either Cartesian coordinates or cylindrical coordinates to 
describe the system, the former being certainly more convenient to express tensor 
fields components while the latter are fit to solve the Dirac equation, as we will 
see. The most important issue in searching for a solution of this problem is the 
choice of appropriate boundary conditions, not an easy task because the Dirac 
equation is a first-order partial differential equation.
The authors of the bag model \cite{bag} have shown that the following condition
\footnote{Actually, in the paper \cite{bag}, the boundary condition chosen is 
$\ii \slashed n \Psi(R) = \Psi (R)$, but the change of sign is indeed immaterial.}
\begin{equation}\label{bagbound}
  \ii \slashed n \Psi(R) = \ii n^\mu \gamma_\mu \Psi(R) = - \Psi (R)  
\end{equation}
ensures the vanishing of the fluxes in eq.~(\ref{flux}) through the border and 
allows non-trivial solutions of the Dirac equation within the cylinder (see 
fig.~\ref{illu}) which, however, extend to the whole space without any discontinuity 
in the field. These boundary conditions, whence the whole solution, are not affected 
by the rotation of a possible material support defining the outer surface of the
cylinder, as, for the inertial observer, the motion transforms the boundary into 
itself. The above equation, in the non-relativistic limit, entails the vanishing 
of the "large" components of the Dirac field at the boundary, that is one is left 
with the Schr\"odinger equation with Dirichlet boundary conditions. The eq.~(\ref{bagbound})
implies that $\Psibar(R)\Psi(R) = 0$ at the boundary \cite{bag}, whence the vanishing
of the outward current flux $j^\mu(R) n_\mu$. Morevoer, since $\Psibar\Psi(R)=0$, 
for any value of $\varphi,z,t$, the outer surface of the cylinder must be such that 
$\partial^\mu \Psibar\Psi |_R = \widehat \Xi(R) \, n^\mu$ or:
\begin{equation}\label{bagbound2}
  n^\mu \partial_\mu (\Psibar\Psi)(R) = \frac{\partial}{\partial r} 
  \Psibar\Psi \Big|_{r=R} = - \widehat \Xi(R)
\end{equation}
Thus, the flux of energy-momentum of the canonical tensor at the boundary is vanishing 
because, using eq.~(\ref{canonical}) and eqs.~(\ref{bagbound}),(\ref{bagbound2}):
\begin{equation*}
 \int_{\partial V} \!\!\!\! \di S \;  \wT^{\mu \nu} n_\mu = 
 \frac {\ii}{2} \int_{\partial V} \!\!\!\! \di S \; \Psibar \slashed n \partial^\nu \Psi 
  - \partial^\nu \Psibar \slashed n \Psi =
   \frac {1}{2} \int_{\partial V} \!\!\!\! \di S \; \Psibar \partial^\nu \Psi + 
  \partial^\nu \Psibar \Psi = \frac {1}{2} \int_{\partial V}\!\!\!\! \di S \; 
  \partial^\nu (\Psibar \Psi) = -\frac {\widehat \Xi (R)}{2} \int_{\partial V} 
  \!\!\!\! \di S \; n^\nu = 0 
\end{equation*}  
Likewise, for the orbital part of the angular momentum flux:
\begin{equation*}
 \int_{\partial V} \di S \;  x^\mu \wT^{\lambda \nu} n_\lambda - (\mu \leftrightarrow \nu)
  = \frac {1}{2} \int_{\partial V} \di S \; x^\mu \partial^\nu (\Psibar \Psi)
  - (\mu \leftrightarrow \nu) = \frac {\widehat \Xi(R)}{2} 
   \int_{\partial V} \di S \; (x^\mu n^\nu -
  x^\nu n^\mu) = 0
\end{equation*}  
where the last integral vanishes because of the geometrical symmetry $z \to -z$.
Finally, the flux of the spin tensor also vanishes at the boundary because, 
using (\ref{canonical}) and (\ref{bagbound}): 
\begin{equation}\label{nspin}
  n_\lambda \wspt^{\lambda, \mu \nu}(R) = 
  \frac{1}{2} \left( \Psibar \slashed n \Sigma^{\mu \nu} 
 \Psi + \Psibar \Sigma^{\mu \nu} \slashed n \Psi \right) =
 -\frac{\ii}{2} \left( \Psibar \Sigma^{\mu \nu} \Psi - \Psibar \Sigma^{\mu \nu} \Psi 
 \right) = 0
\end{equation}
Therefore, the eq.~(\ref{flux}) applies and the integrals:
\begin{equation}\label{generators2}
  \widehat P^\nu =  \int_V \di^3 \x \; \wT^{0 \nu}  \qquad \qquad 
  \widehat J^{\mu \nu} = \int_V \di^3 \x \; \widehat{\cal J}^{0,\mu \nu} 
\end{equation}
are conserved. Since, we also have, from the Lagrangian, the usual anticommutation 
relations at equal times:
\begin{equation*}
  \{ \Psi_a(t,{\bf x}),\Psi^\dagger_b(t,{\bf x}') \} = \delta_{ab} 
  \delta^3({\bf x}-{\bf x}') \qquad 
  \{ \Psi_a(t,{\bf x}),\Psi_b(t,{\bf x}') \} = 
  \{ \Psi^\dagger_a(t,{\bf x}),\Psi^\dagger_b(t,{\bf x}') \} = 0
\end{equation*}    
it is easy to check that the conserved hamiltonian $\ii/2 \int \di^3 \x \; \Psi^\dagger 
\codev_t \Psi$ is indeed, as expected, the generator of time translations, i.e.:
\begin{equation*}		
    [\widehat H, \Psi] = - \ii \frac{\partial}{\partial t} \Psi \qquad
    [\widehat H, \Psi^\dagger] = - \ii \frac{\partial}{\partial t} \Psi^\dagger
\end{equation*}
and, therefore, putting together the above equation with eqs.~(\ref{generators2})
and (\ref{canonical}) we conclude that:
\begin{equation}\label{rotinv}
  [\widehat H, \widehat J_i] = 0
\end{equation}
for the case under examination.     

\begin{center}
\line(1,0){100}
\end{center}

The complete solution of the free Dirac equation for a massive particle in a 
longitudinally unlimited cylinder with finite transverse radius, with boundary 
conditions of the kind (\ref{bagbound}) has been obtained by Bezerra de Mello {\it et
al} in ref.~\cite{masters} and we summarize it here. In a longitudinally unlimited
cylinder, but with finite transverse radius $R$, the field is expanded in terms 
of eigenfunctions of the 
longitudinal momentum, third component of angular momentum, transverse momentum and 
an additional ``spin" quantum number \cite{masters}. The relevant quantum numbers 
${\bf n}= (p_z,M, \zet, \xi)$ take on continuous ($p_z$) and discrete values 
$(M,\zet,\xi)$. 
The third component of the angular momentum $M$ takes on all semi-integer values 
$\pm 1/2,\pm 3/2,\ldots$; the ``spin" quantum number $\xi$ can be $\pm 1$ and the 
transverse momentum quantum number:
\begin{equation}\label{zeta}
  \zet = p_{Tl} R
\end{equation}
takes on discrete values which are zeroes, sorted in ascending order with the label 
$l=1,2,\ldots$ and depending on $M$ and $\xi$, of the equation:
\begin{equation}\label{zero}
 \bsbu{p_T R}\; +\; \sgnm \, \bpms{}{+}\, \bsbd{p_T R} = 0
\end{equation}
where $J$ are Bessel functions and:
\begin{equation}\label{bixi}
  \bpms{}{\pm} = \frac{ \pm m +\xi m_T }{p_T}.
\end{equation}
$m$ being the mass and:
\begin{equation*}
   m_T = \sqrt{p^2_T + m^2}
\end{equation*}
the transverse mass \footnote{In the rest of this section the symbol $p_{Tl}$ stands 
for a discrete variable taking on $(M,\xi,l)$-dependent values given by the 
eq.~(\ref{zeta}) or, later on, by eq.~(\ref{zeta2}).}; we note in passing that 
$\bpms{}{+} = 1/\bpms{}{-}$. The Dirac field itself can be written as an expansion:
\begin{equation}\label{expans}
  \Psi(x) = \displaystyle \sum_\n  U_\n(x) a_\n + V_\n(x) b^\dagger_\n 
\end{equation}
where $a_\n$ and $b_\n$ are destruction operators of quanta $\n$ while:
\begin{equation*}
 \sum_\n \equiv \sum_M \sum_{\xi=\pm 1} \sum_{\zet} \int_{-\infty}^{+\infty} 
 \di p_z = \sum_M \sum_{\xi=-1,1} \sum_{l=1}^\infty \int_{-\infty}^{+\infty} 
 \di p_z
\end{equation*}
The eigenspinors $U_\n$ and $V_\n$ read, in the Dirac representation of the $\gamma$
matrices and in cylindrical coordinates $(t,r,\varphi,z)$:
\begin{eqnarray}\label{spinors}
 &&  U_\n(x) = C_\n \left( \begin{matrix} \besselbetauno{r}\\
                   \ii\, \sgnm \ks \bpms{}{+} \besselbetadue{r} \esp{i\varphi}\\
                   \ks \besselbetauno{r}\\
                   -\ii\, \sgnm\bpms{}{+} \besselbetadue{r}\esp{\ii\varphi}
		   \end{matrix} \right) \frac{1}{\sqrt{2 \pi}} 
		   \esp{\ii\pq{\pt{M-1/2}\varphi + p_z z -\varepsilon t}}
		   \nonumber \\  \nonumber \\
 &&  V_\n(x) = \frac{C_\n}{\bpms{}{-}}\left(\begin{matrix}
                       \besselbetadue{r}\\
                       \ii\, \sgnm \ks \bpms{}{-} \besselbetauno{r} \esp{i\varphi}\\
                       \ks \besselbetadue{r}\\
                       -\ii\, \sgnm\bpms{}{-} \besselbetauno{r}\esp{i\varphi}
                       \end{matrix} \right) \frac{1}{\sqrt{2 \pi}} 
		      \esp{-\ii\pq{\pt{M+1/2}\varphi + p_z z - \varepsilon t}}	                             
\end{eqnarray}
with:
$$
  \ks = \frac{\varepsilon+\xi\sqrt{\varepsilon^2 - p_z^2}}{p_z}
$$
and $\varepsilon = \sqrt{p_z^2 + p_{Tl}^2 + m^2}$ being the energy. The eigenspinors
(\ref{spinors}) are normalized so as to:
\begin{equation*}
  \int_V \di^3 \x \; \Psi^\dagger \Psi = 
  \sum_{\n} a^\dagger_\n a_\n + b_\n b^\dagger_\n 
\end{equation*}
that is with:
\begin{equation}\label{spinorm}
  \int_V \di^3 \x \; U^\dagger_\n(x) U_{\n'}(x) = 
  \int_V \di^3 \x \; V^\dagger_\n(x) V_{\n'}(x) = \delta_{\n\n'} 
  \qquad \qquad \int_V \di^3 \x \; U^\dagger_\n(x) V_{\n'}(x) = 0
\end{equation}
being $\delta_{\n\n'}=\delta_{MM'}\delta_{\xi\xi'}\delta_{ll'}\delta(p_z - p'_z)$
and the anticommutation relations of creation and destruction operators:
\begin{equation}\label{creadest}
  \{ a_\n, a^\dagger_{\n'} \} = \{ b_\n, b^\dagger_{\n'} \} = \delta_{\n\n'} 
  \qquad \qquad \{ a_\n, b_{\n'} \} =  \{ a^\dagger_\n, b_{\n'} \} = 0
\end{equation}  
The normalization coefficient in (\ref{spinors}) obtained from the condition 
(\ref{spinorm}) reads \cite{masters}:
\begin{equation}\label{coeff}
 \pt{C_\n}^{-2} = 2\pi R^2\besselquadrouno{R}
  \frac{\ks^2 +1}{p_{Tl}^2 R^2}\pt{2R^2 m_{Tl}^2 + 2\xi M R m_{Tl} + m R }
\end{equation} 
%

\subsection{Proving the inequivalence}

For what we have seen so far, from a purely quantum field theoretical point of view, 
the Belinfante tensors (\ref{belinfante}) for the Dirac field in the cylinder could 
be regarded as equivalent to the canonical ones in eq.~(\ref{canonical}) because 
they give, once integrated, the same generators (\ref{generators2}). This happens 
because the condition (\ref{boundary}) is met for $\wPhi=\wspt$ (what follows from 
eq.~(\ref{nspin})) and this implies, taking eq.~(\ref{extranti}) into account, that
all the integrands of (\ref{boundary}) vanish at the boundary. Yet, these two set of 
tensors are thermodynamically inequivalent because, as it will be shown hereafter, 
it turns out that, using eq.~(\ref{sptpara}):
\begin{equation}\label{inequa1}
  \spt^{0,ij} = \frac{1}{2} 
  \tr_V ( \wrho :\! \Psibar \{ \gamma^0, \Sigma^{ij} \} \Psi \!: ) 
  = D(r) \epsilon_{ijk} k^k \ne 0  \Rightarrow D(r) \ne 0
\end{equation}
at some $r \ne R$ (we have used the eq.~(\ref{sptpara})), with $\wrho$ written in
eq.~(\ref{thermj}). This will be enough to conclude that either the energy-momentum 
or the angular momentum densities or both have different values for different sets 
of quantum tensors, as previously discussed. Note that the boundary condition 
(\ref{nspin}) together with the general expression of the mean value of the spin 
tensor (\ref{sptpara0}) implies that $D(R)=0$, i.e. its vanishing at the boundary. 

We can rewrite the inequality (\ref{inequa1}) by taking advantage of the commutation 
relation: 
\begin{equation*}
  [\gamma^\lambda, \Sigma^{\mu \nu}] = \ii g^{\lambda \mu} \gamma^\nu - 
  \ii g^{\lambda \nu} \gamma^\mu
\end{equation*}
implying:
\begin{equation*}
  \spt^{0,ij} = \tr_V [ \wrho :\! \Psi^\dagger \Sigma^{ij} \Psi \!: ] 
  - \ii g^{0i} \tr [ \wrho :\!\Psibar \gamma^\nu \Psi \!: ] 
  + \ii g^{0i} \tr [ \wrho :\!\Psibar \gamma^\mu \Psi \!: ] =
  \tr_V [ \wrho :\! \Psi^\dagger \Sigma^{ij} \Psi \!: ] \ne 0
\end{equation*}
or, equivalently:
\begin{equation}\label{dierre}
  D(r) = \frac{1}{2} \epsilon_{ij3} \spt^{0,ij} = \frac{1}{2} 
  \tr [ \wrho :\! \Psi^\dagger \epsilon_{ij3} 
  \Sigma^{i j} \Psi \!: ] \equiv \tr [ \wrho :\! \Psi^\dagger \Sigma_3 \Psi \!: ]
  \ne 0
\end{equation}
where the indices $i,j$ can only take on the value 1 or 2. In the above equation 
and henceforth, we can take the Heisenberg field operators at some fixed time $t=0$ 
because of the stationarity of density operator $\wrho$. Hence we just need to show 
that:
\begin{equation}\label{inequa5}
  \tr_V [ \wrho :\! \Psi^\dagger (0,{\bf x}) \Sigma_z \Psi (0,{\bf x}) \!:] \ne 0
\end{equation}
with 
\begin{equation}\label{sigmaz}
 \Sigma_z= \frac{1}{2}
 \left( {\begin{array}{cccc}1 & 0 & 0 & 0 \\0 & -1 & 0 & 0 \\0 & 0 & 1 & 0 
  \\0 & 0 & 0 & -1\end{array}} \right),
\end{equation}
for some point ${\bf x}$ within the cylinder and our goal is achieved. 

To calculate the mean value of the spin density in eq.~(\ref{inequa5}), we start 
by observing that (see Appendix B for the proof):
\begin{equation}\label{thermave}
 \tr [ \wrho \; a^\dag_\n a_{\n'}] = \frac{\delta_{\n \n'}}
 {\esp{\pt{\varepsilon - M\omega + \mu}/T} + 1}
 \qquad \tr [\wrho \; b^\dag_\n b_{\n'}] = \frac{\delta_{\n \n'}}
 {\esp{\pt{\varepsilon - M\omega - \mu}/T} + 1} \qquad
 \qquad \tr[ \wrho \; a^\dag_\n b_{\n'}] = \tr[ \wrho \; a_\n b_{\n'}] = 0
\end{equation}
which allows us to work it out by plugging in there the field expansion (\ref{expans}):
\begin{equation}\label{density1}
 \tr_V [ \wrho :\! \Psi^\dagger (0,{\bf x}) \Sigma_z \Psi (0,{\bf x}) \!:] 
 = \sum_\n \frac{1}{ \esp{\pt{\varepsilon - M\omega + \mu}/T} +1}
 [U^\dag_\n(x)\Sigma_zU_\n(x)] - \frac{1}{\esp{\pt{\varepsilon - 
 M\omega -\mu}/T} +1} [V^\dag_\n(x)\Sigma_zV_\n(x)]. 
\end{equation}
where we have taken into account that the normal ordering of fermions is such that 
$:\!b_\n b^\dagger_{\n'}\!: = - b^\dagger_{\n'}b_\n $. 
By using eq.~(\ref{spinors}) and (\ref{sigmaz}):
\begin{eqnarray*}
 U^\dag_\n(x) \Sigma_z U_\n(x) &=& \frac{C_\n^2}{4\pi}\pq{\besselquadrouno{r} \;-\; 
 \ks^2\, {\bpms{}{+}}^2 \besselquadrodue{r} \;+\; \ks^2 \besselquadrouno{r} \;-\; 
 {\bpms{}{+}}^2 \besselquadrodue{r}} = \\
 &=& \frac{C_\n^2}{4\pi} \pq{\besselquadrouno{r} \,-\, {\bpms{}{+}}^2 \besselquadrodue{r}}
 \pt{ 1 + \ks^2} 
\\
\\
V^\dag_\n(x) \Sigma_z V_\n(x) &=& \frac{C_\n^2}{4\pi} { {\bpms{}{-}}^2 }
 \pq{\besselquadrodue{r} \,-\, {\bpms{}{-}}^2 \besselquadrouno{r}}\pt{ 1 + \ks^2} = \\
 &=& \frac{C_\n^2}{4\pi} \pq{-\besselquadrouno{r} \,+\, {\bpms{}{+}}^2 \besselquadrodue{r}}
 \pt{ 1 + \ks^2} = -U^\dag_\n(x) \Sigma_z U_\n(x).
\end{eqnarray*}
hence, by using eqs.~(\ref{coeff}) and (\ref{dierre}), we can rewrite eq.~(\ref{density1}) 
as:
\begin{eqnarray}\label{density2}
  && \!\!\!\! \tr_V [ \wrho :\! \Psi^\dagger (0,{\bf x}) 
  \Sigma_z \Psi (0,{\bf x}) \!: ] = D(r) \nonumber \\ 
  && \!\!\!\! = \sum_M \sum_{\xi=\pm 1}\sum_{l=1}^\infty \int_{-\infty}^\infty 
 \di p_z \left[ \frac{1}{\esp{\pt{\varepsilon - M\omega + \mu}/T} +1} +
 \frac{1}{\esp{\pt{\varepsilon - M\omega - \mu}/T} +1} \right]  
 {\frac{p_{Tl}^2\pq{ \besselquadrouno{r} \,-\, {\bpms{}{+}}^2 
 \besselquadrodue{r}}}{8\pi^2 R \besselquadrouno{R}
 \pt{2R m_{Tl}^2 +2\xi M m_{Tl} + m}}}
\end{eqnarray}
The mean value of the spin tensor is therefore given by the sum of a particle and
an antiparticle term which are equal only for $\mu=0$. As expected, it vanishes for 
$r=R$ in view of the eq.~(\ref{zero}), yet our goal is to show that it is non-vanishing 
at some point {\bf x} not belonging to the boundary. It is worth pointing out that, 
if this is the case, the spin tensor has a {\em macroscopic} value because, as it 
is apparent from (\ref{density2}), it is proportional to the
number density (in phase space) of quanta $1/\exp[(\varepsilon - M\omega \pm \mu)/T 
+1]$. 

It is most convenient to consider a point belonging to the rotation axis, i.e. with 
radial coordinate $r=0$ because Bessel functions of all orders but 0 vanish therein. 
By working out the eq.~(\ref{density2}), it can then be shown that $D(0)=0$ for 
$\omega=0$ and that it is an increasing function of $\omega/T$ thereafter (see Appendix 
C). It thence follows that 
\begin{equation*}
  D(0) \ne 0  \qquad {\rm for} \; \; \omega/T > 0
\end{equation*}
and therefore the inequality (\ref{inequa5}) must be true for small, yet finite, 
values of $\omega/T$ around the rotation axis. We note in passing that for $\omega=0$ 
the whole function $D(r)$ must be vanishing because of symmetry reasons. In fact, if 
$\omega=0$, the density operator (\ref{thermjv}) enjoys an additional symmetry, that 
is the rotation of an angle $\pi$ around any axis orthogonal to the cylinder axis, 
say ${\sf R}_2(\pi)$. This transformation corresponds to flip over the cylinder, which 
leaves the system invariant provided that $\omega=0$, and has the consequence that 
any pseudo-vector field directed along the axis must vanish. 

Moreover, it is easy to show, again by using eq.~(\ref{density2}), that the derivative 
of the function $D(r)$ vanishes in $r=0$ for it is proportional to terms, with $N \ge 0$:
\begin{equation*}
   2 J_N (0) J'_N (0) =  J_N (0) \left( J_{N-1}(0) - J_{N+1}(0) \right) 
\end{equation*}
which all vanish because of the known properties of Bessel functions. Hence, the
mean angular momentum density in $r=0$ differs between canonical and Belinfante
tensors, i.e. rewriting the eq.~(\ref{angmomdens2}) for $r=0$:  
\begin{equation*}
 {\boldsymbol{\cal J}}_{\rm Belinfante}(0) = {\boldsymbol{\cal J}}_{\rm canonical}(0) 
 - D(0) \hat{\bf k}
\end{equation*}
where $D(0)$ is finite for finite $\omega$ and positive. Thus, the Belinfante 
angular momentum density is lower than the canonical one by some finite and 
macroscopic amount.

We point out that, had we used the definition of mean values (\ref{mean}) without 
normal ordering, this conclusion would be unaffected. Indeed, the spin tensor 
(\ref{canonical}) is a bilinear in the fields and therefore the difference between
the two definitions is a VEV of the spin tensor: 
\begin{equation}\label{newmean}
     \tr (\wrho :\! \wspt \!: ) = \tr (\wrho \wspt ) - \bra{0} \wspt \ket{0}
\end{equation}
Vectorial irreducible parts of the spin tensor, such as $\wspt^{0,ij}$, have a 
vanishing VEV if the vacuum is invariant under general rotations. The vacuum of 
the free Dirac field in the cylinder - defined by $a_{\bf n} \ket{0} = 0$ - is 
indeed rotationally invariant. If degenerate vacua existed, the commutation 
of the hamiltonian with angular momentum operators that was shown before (see 
eq.~(\ref{rotinv})) would ensure that they belong to some irreducible representation 
of the SU(2) group. However, for the free Dirac field in the cylinder, the angular 
momentum operator along the $z$ axis turns out to be:
\begin{equation*}
  \widehat J_z = \sum_{\bf n} M (a^\dagger_{\bf n} a_{\bf n} + b^\dagger_{\bf n} 
  b_{\bf n})
\end{equation*}
so that $\widehat J_z \ket{0} = 0$ on all possible degenerate vacua. This means 
that the only possible multiplet is one-dimensional and, thereby, the vacuum is 
non-degenerate and the second term in the eq.~(\ref{newmean}) vanishes.

\subsection{The non-relativistic limit}

It would be very interesting to calculate the function $D(r)$ numerically to ``see"
the difference between the Belinfante and the canonical tensors and to make sure
that this difference is not a rapidly oscillating function on a microscopic scale,
which would render the macroscopic observation of the difference impossible. This 
is, though, very hard in the fully relativistic case but relatively easy in the 
non-relativistic limit $m/T \gg 1$ because in this case the eq.~(\ref{zero}) 
yielding the quantized transverse momenta reduces to the vanishing of one single 
Bessel function. This happens because in the non-relativistic limit:
\begin{equation}\label{nonrel0}
 \bpms{}{+} = \frac{\xi m_T + m}{p_T} = \begin{cases} \frac{m_T + m}{p_T} \simeq 
 \frac{2m + p^2_T/2m}{p_T} \simeq \frac{2m}{p_T} \gg 1 \qquad {\rm for} \; \xi=1 \\
 \frac{m - m_T}{p_T} \simeq \frac{p^2_T/2m}{p_T} \simeq \frac{p_T}{2m} \ll 1 
 \qquad {\rm for} \; \xi=-1 \end{cases}
\end{equation}
so that the eq.~(\ref{zero}) in fact reduces to:
\begin{equation}\label{nonrel1}
  \begin {cases} \bsbd{p_T R} = \sgnm \frac{p_T}{2m} \bsbu{p_T R} \simeq 0  
  \qquad {\rm for} \; \xi=1 \\
  \bsbu{p_T R} = \sgnm \frac{p_T}{2m} \bsbd{p_T R} \simeq 0 
  \qquad {\rm for} \; \xi=-1 \end{cases}
\end{equation}
Altogether, we can solve the equation $J_L(p_T R)=0$ for all integers $L = M +\xi/2$ 
and take the quantized transverse momenta:
\begin{equation}\label{zeta2}
  p_{Tl}=\frac{\zeta_{L,l}}{R}
\end{equation}
where $\zeta_{L,l}\;\; l=1,2,\ldots$ are now the familiar zeroes of the Bessel 
function of integer order $L$. 
 
We can now separate the particle and antiparticle terms in the eq.~(\ref{density2}):
\begin{equation*}
  D(r)^\pm = \sum_M \sum_{\xi=\pm 1} \sum_{l=1}^\infty \int_{-\infty}^\infty 
 \di p_z \frac{1}{\esp{\pt{\varepsilon - M\omega \pm \mu}/T} +1}  
 {\frac{p_{Tl}^2\pq{ \besselquadrouno{r} \,-\, {\bpms{}{+}}^2 
 \besselquadrodue{r}}}{8\pi^2 R \besselquadrouno{R}
 \pt{2R m_{Tl}^2 + 2\xi M m_{Tl} +m}}}
\end{equation*}
with $D(r)=D(r)^++D(r)^-$. In the non-relativistic limit one has:
\begin{equation}\label{appro1}
 2R m_T^2 + 2\xi M m_T +m \simeq 2 R m^2 + 2\xi M m + m \simeq 2 R m^2
\end{equation}
where the last approximation is due to the obvious assumption $Rm \gg 1$ and that 
the term $|\xi M m|$ can be comparable to $Rm^2$ only if $|M|$ is very large. However, 
terms with large $|M|$ are either suppressed by the exponential $\exp[\omega M/T]$ 
or by the Bessel functions, which effectively implements the semiclassical equality 
$M \approx R p_T$; since non-relativistically $R m^2 \gg R p_T m \approx |M| m$, the 
approximation (\ref{appro1}) is justified. We then calculate the terms with $\xi=1$
and $\xi=-1$ in the sum in eq.~(\ref{nonrel1}) separately. For $\xi=1$ one sets 
$M + 1/2 = L$ and writes the integrand of eq.~(\ref{density2}), including approximation 
(\ref{appro1}) and taking into account (\ref{nonrel0}):
\begin{equation}\label{appro2}
 \frac{1}{\esp{\pt{\varepsilon - L\omega + \omega/2 \pm \mu}/T} +1}
 \frac{p^2_T}{16 \pi^2 R^2 m^2}\frac{J^2_{|L-1|}(p_{Tl} r) \,-\, \frac{4 m^2}{p_{Tl}^2}
 J^2_{|L|}(p_{Tl} r)}{J^2_{|L-1|}(p_{Tl} R)} \simeq -
  \frac{1}{\esp{\pt{\varepsilon - L\omega + \omega/2 \pm \mu}/T} +1}
 \frac{1}{4 \pi^2 R^2}\frac{J^2_{|L|}(p_{Tl} r)}{J^2_{|L-1|}(p_{Tl} R)}
\end{equation} 
where $p_{Tl}$ is a solution of the first equation in (\ref{nonrel1}). Similarly, for 
$\xi=-1$ one sets $M -1/2 = L$ and obtains, by using the second of the equations 
(\ref{nonrel1}):
\begin{eqnarray}\label{appro3}
 && \frac{1}{\esp{\pt{\varepsilon - L\omega + \omega/2 \pm \mu}/T} +1}
 \frac{p^2_T}{16 \pi^2 R^2 m^2}\frac{J^2_{|L|}(p_{Tl} r) \,-\, \frac{p_{Tl}^2}{4 m^2}
 J^2_{|L+1|}(p_{Tl} r)}{J^2_{|L|}(p_{Tl} R)} \simeq 
  \frac{1}{\esp{\pt{\varepsilon - L\omega - \omega/2 \pm \mu}/T} +1}
\frac{p^2_T}{16 \pi^2 R^2 m^2} 
\frac{J^2_{|L|}(p_{Tl} r)}{\frac{p_{Tl}^2}{4m^2}J^2_{|L+1|}(p_{Tl} R)} \nonumber \\
&&= \frac{1}{\esp{\pt{\varepsilon - L\omega - \omega/2 \pm \mu}/T} +1}
\frac{1}{4 \pi^2 R^2}\frac{J^2_{|L|}(p_{Tl} r)}{J^2_{|L+1|}(p_{Tl} R)} 
\end{eqnarray} 
Now, by using approximations (\ref{appro1}),(\ref{appro2}) and (\ref{appro3})
we can write the non-relativistic limit of $D(r)^\pm$ as:
\begin{equation}\label{nonrel3}
 D(r)^\pm = \frac{1}{4\pi^2 R^2} \sum_{L=-\infty}^\infty \sum_{l=1}^\infty  
 \int_{-\infty}^\infty \di p_z  \frac{1}{\esp{\pt{\varepsilon - L\omega - \omega/2 
  \pm \mu}/T} +1} \frac{J^2_{|L|}(p_{Tl} r)}{J^2_{|L+1|}(p_{Tl} R)} - 
  \frac{1}{\esp{\pt{\varepsilon - L\omega + \omega/2 \pm \mu}/T}+1} 
  \frac{J^2_{|L|}(p_{Tl} r)}{J^2_{|L-1|} (p_{Tl} R)} 
\end{equation}
where the first term is to be associated to particles with spin projection $+1/2$
along the $z$ axis and the second term to those with projection $-1/2$. Finally, 
the integral over $p_z$ in eq.~(\ref{nonrel3}) can be worked out by first introducing 
the non-relativistic approximation $\varepsilon = m + p_T^2/2m + p_z^2/2m$ and 
then expanding the Fermi distribution. The final result is:
\begin{eqnarray*}\label{nonrel4}
 D(r)^\pm &=& \frac{1}{4\pi^2 R^2} \sum_{L=-\infty}^\infty \sum_{l=1}^\infty
  \sum_{n=1}^\infty (-1)^{n+1} \sqrt{\frac{2 \pi m KT}{n}} \e^{-n 
  (mc^2 \pm \mu + p^2_T/2m - L \hbar \omega)/KT}  \nonumber \\ 
  &\times& \left\{ \e^{ n \hbar\omega/2KT}
  \frac{J^2_{|L|}(p_{Tl} r/\hbar)}{J^2_{|L+1|}(p_{Tl} R/\hbar)} - \e^{- n \hbar\omega/2KT} 
  \frac{J^2_{|L|}(p_{Tl} r/\hbar)}{J^2_{|L-1|} (p_{Tl} R/\hbar)} \right\}
\end{eqnarray*} 
where we have purposely restored, for reasons to become clear shortly, the natural
constants. 

It is very interesting to observe that the functions $D(r)^\pm$, hence $D(r)$, are 
non-vanishing in the exact non-relativistic limit $c \to \infty$. Indeed, it can be 
seen from eq.~(\ref{nonrel4}) that no factor $\hbar$ or $c$ or powers thereof appear 
as proportionality constants in front of it, because the $D(r)^\pm$ dimension is already 
- in natural units - that of an angular momentum; the only $c^2$ needed is in the 
exponent, which is compensated by a shift of the chemical potential, 
and the only $\hbar$'s needed are those multiplying $\omega$ and in the argument 
of Bessel functions. Since $\hbar$ multiplies $\omega$ everywhere and $D(r)$ 
vanishes for $\omega=0$, we also see that the difference between canonical and 
Belinfante densities is essentially a quantum effect, as it vanishes in the limit 
$\hbar \to 0$; this is expected as the spin tensor exists only for quantum fields.
  
For very small values of $\hbar\omega/KT$ these two functions are proportional to 
$\hbar\omega/KT$ itself since $D(r)|_{\omega=0} = 0$, as discussed in the previous
subsection.  
Retaining only the $n=1$ term of the series, corresponding to the Boltzmann limit
of Fermi-Dirac statistics, and expanding the exponentials $\exp(\pm n\hbar \omega/2KT)$
at first order, one obtains the noteworthy equality:
\begin{equation}\label{meaning}
  D(r)^\pm = \hbar \tr [ \wrho (:\!\Psi^\dagger \Sigma_z \Psi \!:)^\pm] 
  \simeq \frac{1}{2} \frac{\hbar \omega}{KT} \hbar 
   \tr [ \wrho (:\!\Psi^\dagger \Psi \!:)^\pm]
  = \hbar \frac{1}{2} \frac{\hbar \omega}{KT} \left(\frac{\di n}{\di^3 \x}\right)^\pm
\end{equation}
where the superscript $\pm$ implies that one retains either the particle or the
antiparticle term in the expansion of the free field and $(\di n/\di^3 \x)^\pm$ is,
apparently, the particle or antiparticle density. The eq.~(\ref{meaning}) can be
shown by retracing all the steps of the calculations carried out for the spin 
tensor just replacing $\Sigma_z$ with the identity matrix.

\begin{figure}[ht]
\begin{minipage}[b]{0.4\linewidth}
\centering
\includegraphics[scale=0.4]{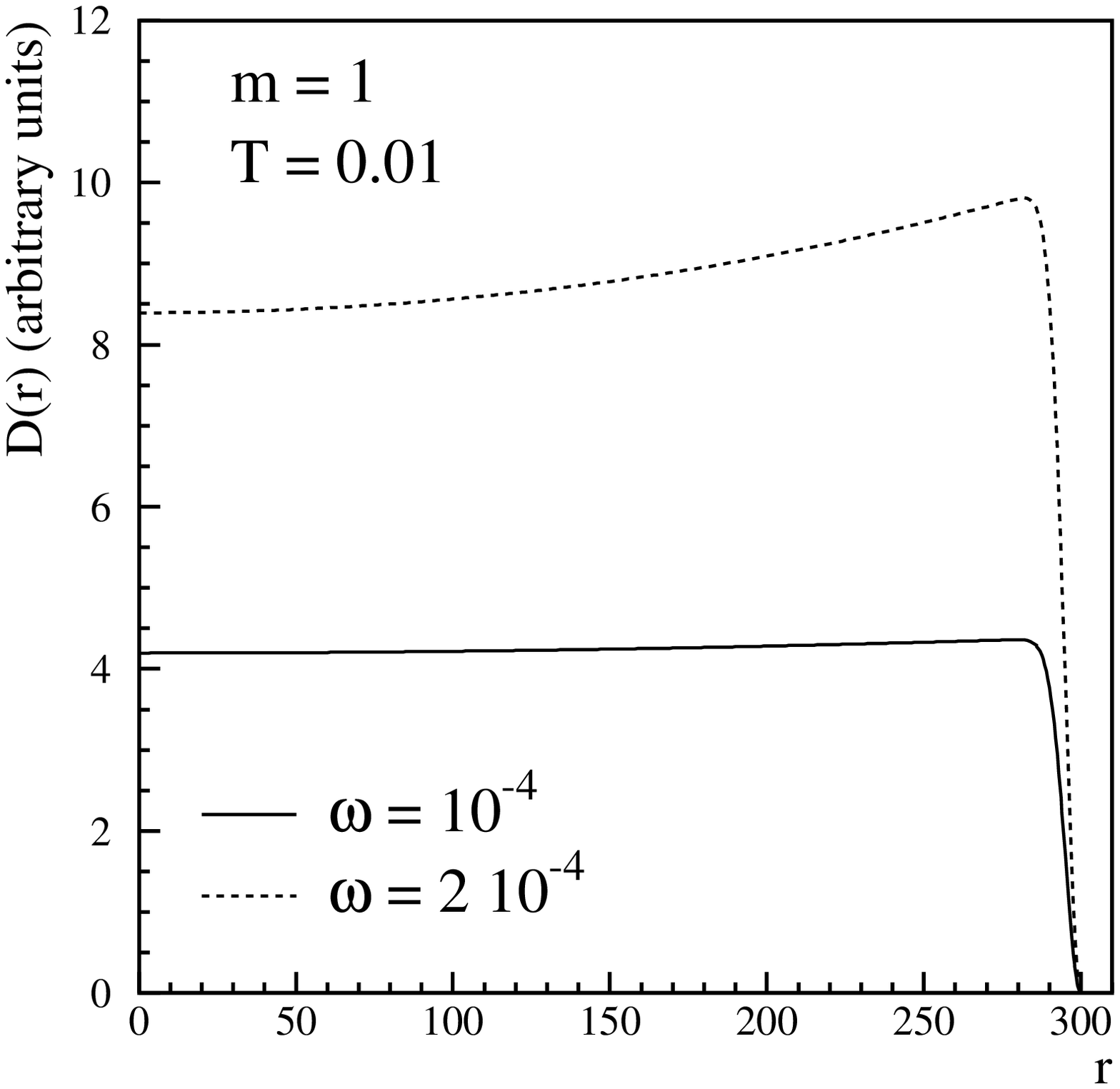}
\caption{Function $D(r)$, corresponding to the mean value of the canonical spin 
tensor for the free Dirac field, in a rotating cylinder at thermodynamical equilibrium
as a function of radius $r$, in the non-relativistic limit.}
\label{spin1}
\end{minipage}
\hspace{0.5cm}
\begin{minipage}[b]{0.4\linewidth}
\centering
\includegraphics[scale=0.4]{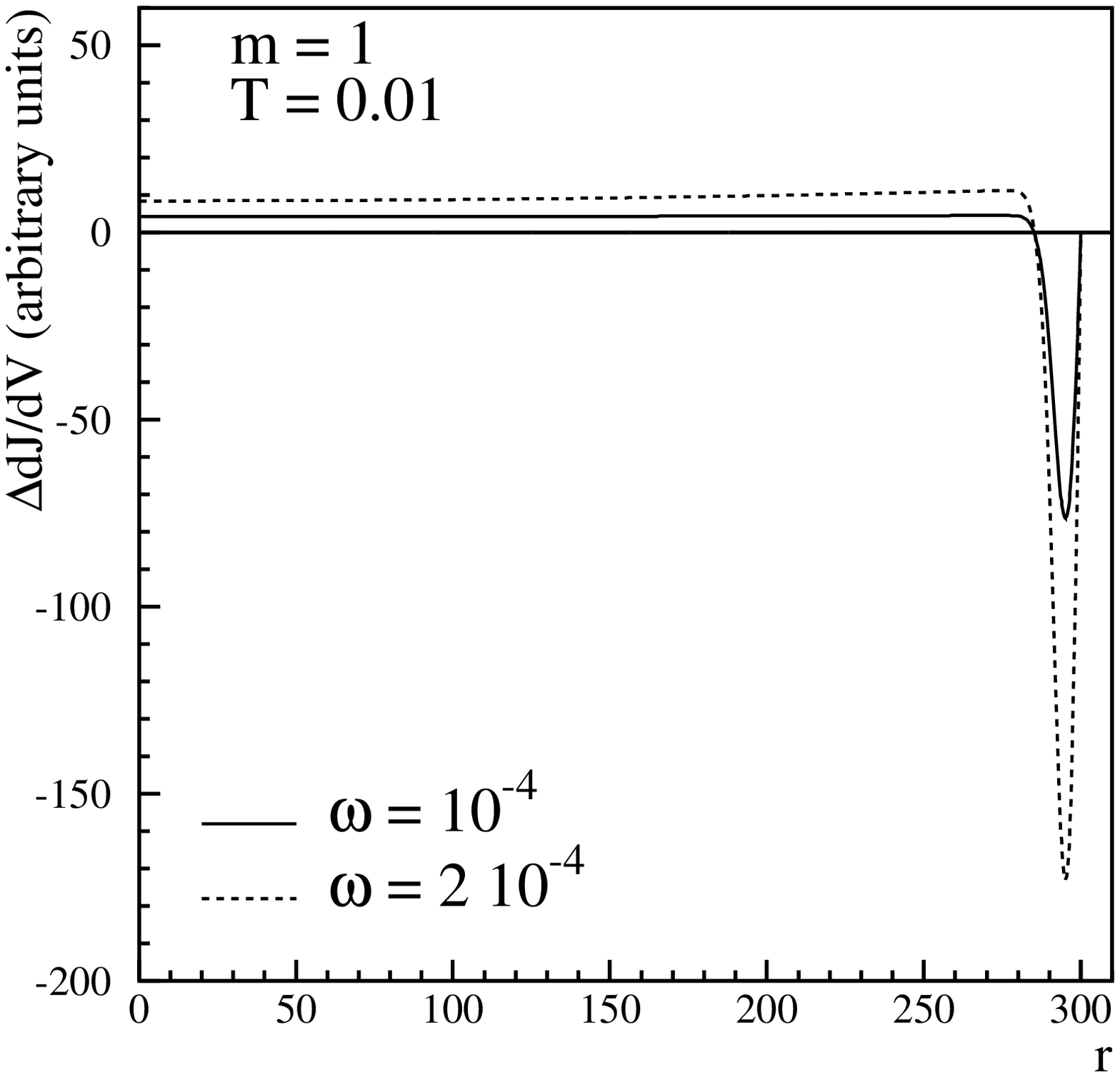}
\caption{Difference between the mean value of the canonical angular momentum density 
and the Belinfante angular momentum density for the free Dirac field, in a rotating 
cylinder at thermodynamical equilibrium as a function of radius $r$, in the 
non-relativistic limit.}
\label{spin2}
\end{minipage}
\end{figure}

The function $D(r)$ can be computed with available numerical routines finding a 
sufficient number of zeroes of Bessel functions, according to eq.~(\ref{nonrel1}). 
For the numerical computation to be accurate enough one has to make the series in 
$L$, $l$ and $n$ quickly convergent at any $r$. For the series in $L$, two requirements
should be met: first (in natural units) $\omega/T \ll 1$ in order to keep the 
exponential $\exp[L \omega/T]$ relatively small and, secondly, the radius $R$ 
should be such that $R \sqrt{mT}$ is not too large; this condition stems from 
the fact that, as the Bessel functions effectively implement the semiclassical 
approximation $|L| \simeq p_T R$ and $p_T \approx \sqrt{mT}$, the effective maximal 
value of $L$ is of the order of $R \sqrt{mT}$. For the series in $l$, one has to 
set $m/T \gg 1$, so that large $p_T$'s are strongly suppressed; this is also the 
non-relativistic limit condition. For the series in $n$, one has to choose $\mu$ 
so as to keep far from the degenerate Fermi gas case. The function $D(r)$ as a 
function of $r$ is shown in fig.~\ref{spin1} for $\mu = 0$, $R=300$, $T=0.01$, 
$m=1$ and two different values of $\omega$, $10^{-4}$ and $2\cdot 10^{-4}$; the 
function $(r/2)D'(r)-D(r)$, which is the difference between angular momentum densities 
for the canonical and Belinfante tensors, is shown in fig.~\ref{spin2}. 
  
The plots in figs.~\ref{spin1},\ref{spin2} show that the angular momentum density 
is larger in the canonical than in the Belinfante case almost everywhere, except 
for a narrow space near the boundary, whose thickness is plausibly determined by 
the microscopic
scales of the problem (thermal wavelength or Compton wavelenght). Thereby, the 
observable macroscopic value of the differences between angular momentum densities, 
for a rotating system of free fermions, is the slowly varying positive one in the 
bulk. While the boundary conditions are needed to ensure the invariance of the 
total angular momentum, the rapid drop to zero within a microscopic distance from 
the cylinder surface tells us that the chosen boundary conditions at a mascroscopic
scale of observation correspond to a discontintuity or a surface effect. Any 
{\em macroscopic} coaxial sub-cylinder of the full cylinder with a radius $r<R$ will 
therefore have different total angular momenta whether one chooses the canonical or 
the Belinfante tensors in eqs.~(\ref{canonical}) and (\ref{belinfante}) respectively. 
Such an ambiguity is physically unacceptable and can be solved only by admitting 
that these tensors are in fact inequivalent.

\section{Conclusions and outlook}
\label{concl}

In conclusion, we have shown that, in general, couples of stress-energy and spin 
tensors related by a pseudo-gauge transformation (\ref{transfq}) and allegedly 
equivalent in quantum field theory, are in fact thermodynamically inequivalent.
The inequivalence shows up only for thermodynamical rotating systems and not for 
the systems - familiar in thermal field theory - locally at rest in an inertial 
frame. We have worked out exhaustively an instance of such inequivalence 
involving the free Dirac field and shown that, surprisingly, the canonical and 
Belinfante tensors imply the same mean energy density but different mean densities 
of momentum and angular momentum. Particularly, the latter is almost everywhere 
larger in the canonical than in the Belinfante case for a small, yet macroscopic, 
amount. We would like to stress that this result does not to depend on an 
inappropriate treatment of the quantum field problem in a region with finite 
transverse size. First of all, field boundary conditions have been chosen so as 
to guarantee the invariance of the integrated quantities - i.e. the generators - 
and, secondly, the final spin tensor value (\ref{meaning}) is proportional to 
particle density through fixed parameters $\omega$ and $T$. Hence, its non-vanishing 
does not apparently depend on spurious factors depending on the radius $R$ which 
would probably be there if that was the result of an inaccurate calculation of 
finite size effects. 

What is the right couple of tensors? Needless to say, this is a very important issue; 
for instance, if it was found that the quantum spin tensor is not the trivial Belinfante
one (i.e. vanishing) this would have major consequences in hydrodynamics and gravity, 
even more if its associated stress-energy tensor had a non-symmetric part, because
this could imply a torsion of the spacetime (for a recent discussion see e.g. 
ref.~\cite{bjorken}). Thus far in this work, no method has been discussed to 
answer the above question. Indeed, the problem of determining the right tensors 
can be approached both theoretically and experimentally.

From an experimental viewpoint, in principle we could decide if a specific stress-energy 
or a spin tensor is {\em wrong} by measuring with sufficient accuracy the angular 
momentum density of a rotating system at full thermodynamical equilibrium kept at 
fixed temperature $T$ and angular velocity $\omega$. This measurement would, for 
instance, be able to reject the canonical or Belinfante tensor without even the need
of resorting to relativistic systems as their difference has a non-vanishing 
non-relativistic limit, as has been discussed at the end of last section. 
In practice, at a glance, this measurement would not seem an easy one. According to
eq.~(\ref{meaning}), in the non-relativistic limit the difference between these 
two tensors is of the order of $\hbar\omega/KT$ times $\hbar$ times the particle 
density, that is particles have a polarization of the order of $\hbar\omega/KT$.
This ratio is extremely small for ordinary macroscopic systems; assuming a large 
angular velocity $\omega$, say 100 Hz, at room temperature $T=300\, {\rm ^o K}$
it turns out to be of the order of $10^{-12}$. Notwithstanding, this is precisely 
the polarization responsible for the observed magneto-mechanical phenomena, the Barnett 
\cite{barnett} (magnetization induced by a rotation) and Einstein-De Haas (rotation 
induced by magnetization) effects. It is therefore possible that with some suitable 
experiment of this sort one can discriminate between spin tensors; this will be 
the subject of further investigation. We would like to point out that the effect 
could be enhanced lowering the temperature so much to increase the ratio 
$\hbar \omega/KT$, e.g. with cold atom techniques.

From a theoretical viewpoint, we cannot, for the present, determine a thermodynamically
``best" couple of stress-energy and spin tensor. Yet, we can argue, on the basis of
a thermodynamical argument - which can in principle be used to assess any other 
couple of tensors - that the canonical spin tensor is favoured over the Belinfante 
one; the argument also elucidates why the source of the inequivalence is ultimately 
the second law of thermodynamics. Let us first write down the entropy of a system
with cylindrical symmetry and large but finite radius $R$, using eq.~(\ref{thermj}):
\begin{equation*}
  S = -\tr_V [\wrho \log \wrho] = \log Z_\omega + 
  \frac{\langle \widehat H \rangle}{T} - \frac{\mu}{T} \langle \widehat Q \rangle -
  \frac{\omegav}{T} \cdot \langle \widehat {\bf J} \rangle
\end{equation*}
where $\langle \;\; \rangle$ stands for $\tr [\wrho \;\;]$. All average quantities
in the above equation, namely energy, charge and angular momentum, have a density, 
meaning that they are given by a volume integral of a supposedly objective function.
Now let us assume that also entropy has a physically objective density, what is 
expected to happen for a macroscopic system; in this case, according to the above 
equation, the potential $\log Z_\omega$ has an objective density as well. Furthermore,
if the system is properly thermodynamical, the derivatives of this density 
$\di \log Z_\omega/\di V$ with respect to the intensive parameters $1/T$,$\mu/T$ and 
$\omega/T$ should give the energy density, the charge density and the angular 
momentum density, similarly to what happens for the integrated quantities. 

Suppose that the system is initially non rotating, i.e. with $\omega=0$. We have 
seen in Sect.~\ref{dirac} that in this case the function $D(r)$ vanishes, hence 
there is no difference between the mean values of canonical and Belinfante tensors. 
Let us then focus on a coaxial sub-cylinder with radius $r \ll R$, yet large 
enough for it to be macroscopic. Keeping the temperature $T$ and chemical potential 
$\mu$ fixed, let us turn on very slowly a small angular velocity $\Delta \omega$. 
Thermodynamically, we can think of the sub-cylinder as ``the system" and the rest 
as an angular momentum reservoir at full thermodynamical equilibrium with it as 
both have the same angular velocity. It is well known that a thermodynamical system 
at fixed $T$, $V$ and $\mu$ maximizes the value of the thermodynamical potential 
$\log Z$. Likewise, a thermodynamical system with fixed $T$, $V$, $\mu$ and 
$\omega$ maximizes the value of the potential $\log Z_\omega$. For the sub-cylinder
with radius $r$, according to our previous thermodynamical assumptions, this potential 
is given by the integral over the region $V_r$ of the density $\di \log Z_\omega/\di V$ 
and its variation after switching on the rotational motion is:
\begin{equation*}
   \Delta \log Z_{\omega,r} \simeq 
   \frac{\partial \log Z_{\omega,r}}{\partial(\omega/T)}\Bigg|_{V_r,T,\mu}
   \Delta \left(\frac{\omega}{T}\right) = \langle J \rangle_r 
   \Delta \left(\frac{\omega}{T}\right)  
\end{equation*}
where $\langle J \rangle_r$ is the angular momentum of the sub-cylinder. This
variation should be maximal and since, for what we have seen in Sect.~\ref{dirac}, 
in any properly macroscopic coaxial sub-cylinder with radius $r<R$, the canonical 
value of the angular momentum is larger than the Belinfante one, the final value 
of the thermodynamical potential of the sub-cylinder will be larger in the canonical 
than in the Belinfante case. Hence, if the system could choose between canonical 
and Belinfante tensors, the former would be certainly favoured. Of course, this 
method allows to discriminate between two couples of tensors but, for the present, 
does not permit to single out the ``best" couple of tensors for a given quantum 
field theory. For instance, for the free Dirac field, it may happen that neither 
the canonical nor the Belinfante tensors maximize the thermodynamic potential of 
such a subsystem and another couple does.

\section*{Acknowledgements}

We are greteful to A.~Cappelli, F.~Colomo, M.~Lenti, M.~Redi, D.~Seminara 
for useful discussions and revision of the manuscript.



\appendix

\section*{APPENDIX A - Action and canonical stress-energy tensor for the Dirac 
field in a bounded region}

Consider an action $A$ of general fields $\psi^a$ in a cylindrical region in
fig.~\ref{illu} and its variations:
\begin{eqnarray}\label{daction}
  \delta A &=& \int_V \di^4 x \; \left( \frac{\partial{\cal L}}{\partial \psi^a}
  - \partial_\mu \frac{\partial{\cal L}}{\partial \partial_\mu \psi^a} \right)
  \delta \psi^a + \int_V \di^4 x \; \partial_\mu \left( 
  \frac{\partial{\cal L}}{\partial \partial_\mu \psi^a} \delta \psi^a \right)
  \nonumber \\
  &=& \int_V \di^4 x \; \left( \frac{\partial{\cal L}}{\partial \psi^a}
  - \partial_\mu \frac{\partial{\cal L}}{\partial \partial_\mu \psi^a} \right)
  \delta \psi^a + \int_{\partial V} \di t \di S \; n^\mu \frac{\partial{\cal L}}
  {\partial \partial_\mu \psi^a} \delta \psi^a 
\end{eqnarray}
Particularly, if we take the Dirac action:
\begin{equation}\label{action}
  A = \int_V \di^4 x \; 
  \left( \frac{\ii}{2} \Psibar \gamma^\mu \codev_\mu \Psi - m \Psibar \Psi
  \right)
\end{equation}
and require it to be stationary with respect to variation of the fields $\Psi,\Psibar$ 
with the boundary conditions (\ref{bagbound}), we obtain the free Dirac equation.  
This can be shown by working out the derivatives in the boundary integral with 
$\cal L$ as in eq.~(\ref{lagra}):
\begin{equation*}
  n^\mu \frac{\partial{\cal L}} {\partial \partial_\mu \psi^a} \delta \psi^a 
 = n^\mu \frac{\ii}{2} \Psibar \gamma_\mu \delta \Psi - n^\mu \frac{\ii}{2} 
 \delta \Psibar \gamma_\mu \Psi =  \frac{\ii}{2} (\Psibar \slashed n \delta \Psi
 - \delta \Psibar \slashed n \Psi )
\end{equation*}
As the fields meet boundary condition (\ref{bagbound}):
\begin{equation*}
\frac{\ii}{2} (\Psibar \slashed n \delta \Psi - \delta \Psibar \slashed n \Psi)
= -\frac{1}{2} (\Psibar \delta \Psi + \delta \Psibar \Psi) = -\frac{1}{2} 
 \delta (\Psibar \Psi)
\end{equation*}
as well as $\Psibar \Psi (R) = 0$ as a consequence of (\ref{bagbound}), then 
$\delta \Psibar \Psi$ vanishes at the boundary and the second integral in 
eq.~(\ref{daction}) vanishes. We are thus left with a bulk integral which has 
to vanish for general variations of the field, leading to free Dirac equations, 
as usual.
 
For the calculation of canonical tensors from the action (\ref{action}), we follow 
ref.~\cite{weinberg} and use a space-dependent variation of the fields:
\begin{equation*}
  \delta \Psi(x) = \Psi(x + \epsilon(x)) - \Psi(x) \simeq 
   \epsilon^\mu(x) \partial_\mu \Psi(x)
\end{equation*}
with small $\epsilon(x)$. This is a particular variation of the field which 
fulfills the condition (\ref{bagbound}) if $\epsilon(x)=0$ at the boundary and
this is what we set. If $\Psi$ is the solution of the equation of motion, then 
the variation of the action should vanish. After some easy calculations (see 
ref.~\cite{weinberg}):
\begin{equation*}
 0 = \delta A = \int_V \di^4 x \; \partial_\mu \left( 
 \frac{\partial{\cal L}}{\partial \partial_\mu \psi^a} \partial^\nu \psi^a
 - g^{\mu\nu} {\cal L} \right) \epsilon_\nu + 
 \int_V \di^4 x \; \partial_\mu \left( 
 \frac{\partial{\cal L}}{\partial \partial_\mu \psi^a} \partial^\nu \psi^a
 \epsilon_\nu \right)
\end{equation*}
The second term can be turned into a boundary integral which vanishes because
$\epsilon=0$ there, as has been mentioned. The first term should also vanish 
and since $\epsilon(x)$ is an arbitrary function, the divergence of the what
can be easily recognized as the canonical stress-energy tensor must vanish.

A similar reasoning leads to the conclusion that the angular momentum tensor is 
conserved.

\section*{APPENDIX B - Calculation of the mean value of products of creation
and destruction operators}

We follow the argument used in \cite{degroot}. The aim is to calculate:
\begin{equation*}
 \tr_V [\wrho \, a^\dagger_\n a_{\n'}]
\end{equation*}
with $\wrho$ given by eq.~(\ref{thermj}). For this purpose we define, with $\beta=1/T$:
\begin{equation}\label{abeta}
  a^\dagger_\n(\beta) = \e^{-\beta(\widehat H - \omega \widehat J - \mu \widehat Q)} 
  a^\dagger_\n \e^{\beta(\widehat H - \omega \widehat J - \mu \widehat Q)}
\end{equation}
and similarly for $a_\n$, $b_\n$ and $b^\dagger_\n$. From the above equation it
ensues:
\begin{equation}\label{differential1}
  \frac{\partial a^\dagger_\n(\beta)}{\partial \beta} = [a^\dagger_\n(\beta),
  \widehat H - \omega \widehat J -\mu \widehat Q]
\end{equation}
and, since:
\begin{equation*}
   [\widehat H, a^\dagger_\n] = \varepsilon a^\dagger_\n \qquad 
   [\widehat J, a^\dagger_\n] = M a^\dagger_\n  \qquad 
   [\widehat Q, a^\dagger_\n] = q a^\dagger_\n 
\end{equation*}
one readily obtains that eq.~(\ref{differential1}) is equivalent to:
\begin{equation*}
  \frac{\partial a^\dagger_\n(\beta)}{\partial \beta} = 
  (-\varepsilon + M \omega + \mu q) a^\dagger_\n(\beta)
\end{equation*}
which is solved by, being $a^\dagger_\n(0)= a^\dagger_\n$ :
\begin{equation}\label{abeta2}
   a^\dagger_\n(\beta) =  
   a^\dagger_\n \e^{-\beta(\varepsilon - M \omega - \mu q)}
\end{equation}

We can now write:
\begin{eqnarray*}
 &&\tr_V [\wrho \, a^\dagger_\n a_{\n'} ] = \tr_V [\wrho \, a^\dagger_\n 
 \e^{\beta(\widehat H - \omega \widehat J -\mu \widehat Q)} 
 \e^{-\beta(\widehat H - \omega \widehat J -\mu \widehat Q)} 
 a_{\n'} ] = \tr_V [\e^{-\beta(\widehat H + \omega \widehat J -\mu \widehat Q)} 
 a_{\n'} \,\wrho \, a^\dagger_\n \e^{\beta(\widehat H - \omega \widehat J -\mu 
 \widehat Q)}] \nonumber \\
 = &&  \frac{1}{Z_\omega} \tr_V [\e^{-\beta(\widehat H - \omega \widehat J - 
 \mu \widehat Q)} a_{\n'} \, \e^{-\beta(\widehat H - \omega \widehat J - 
 \mu \widehat Q)} a^\dagger_\n \e^{\beta(\widehat H - \omega \widehat J - 
 \mu \widehat Q)}] = \frac{1}{Z_\omega} \tr_V [\e^{-\beta(\widehat H - 
 \omega \widehat J -\mu \widehat Q)} 
 a_{\n'} \, a^\dagger_\n(\beta)] =  \tr_V [\wrho \, a_{\n'} a^\dagger_\n(\beta)]
\end{eqnarray*}
where we have used the ciclicity of the trace, the definition of $\wrho$ in 
eq.~(\ref{thermj}) and the eq~(\ref{abeta}). It should be pointed out that 
the ciclicity of the trace can be used safely because a complete set of states 
for the cylinder with finite radius can be constructed with eigenvectors of 
the operators $\widehat H$,$\widehat J_z$ and $\widehat Q$; we could have also
used the full trace and insert the operator $\Pro_V$ discussed in 
Sect.~\ref{thermang} but this would have not changed the final result as this 
operator commutes with $\wrho$. By using eq.~(\ref{abeta2}) and the 
anticommutation relation (\ref{creadest}), the above equation can also be written as:
\begin{equation*}
   \tr_V [\wrho \, a^\dagger_\n a_{\n'} ] = \tr_V [\wrho \, a_{\n'} a^\dagger_\n(\beta)]
= \tr_V [\wrho \, a_{\n'} a^\dagger_\n ] \e^{-\beta(\varepsilon - M \omega -\mu q)}
= \left( - \tr_V [\wrho \,  a^\dagger_\n a_{\n'}] + \delta_{\n \n'} \right)
   \e^{-\beta(\varepsilon - M\omega -\mu q)}
\end{equation*}
whence:
\begin{equation*}
  \tr_V [\wrho \, a^\dagger_\n a_{\n'}] = 
  \frac{\delta_{\n\n'}}{\e^{\beta(\varepsilon - M \omega -\mu q)}+1}
\end{equation*}
The above method can be used for the calculation of other bilinear combinations of 
creation and destruction operators, leading to the equalities reported in 
eq.~(\ref{thermave}).

\section*{APPENDIX C - Calculation of the function $D(r)$ on the rotation axis}

We calculate $D(0)|_{\omega=0}$ to show that it is vanishing as well as its derivative
with respect to $\omega/T$ to show that is strictly positive. The function $D(r)$ in 
eq.~(\ref{density2}) is the sum of a particle $D(r)^+$ and an antiparticle $D(r)^-$ 
term: we focus on the particle term as the calculation for antiparticle is a 
trivial extension. Since Bessel functions of all orders but zero vanish ($J_0(0)=1$), 
in $r=0$ in the sum of eq.~(\ref{density2}) only terms with $M=-1/2$ and $M=1/2$ 
survive:  
\begin{eqnarray}\label{density3}
 D(0)^+ 
 &=& \frac{1}{8\pi^2 R} \sum_{\xi=\pm 1}\sum_{l=1}^\infty \int_{-\infty}^\infty 
 \di p_z \left\lbrace \frac{\ppl^2}{\pt{\esp{\pt{\varepsilon-\frac{1}{2}\omega+\mu}/T}+ 1} 
 J_0(\ppl R)^2 \pq{2R\pt{\ppl^2 + m^2} + \xi\sqrt{\ppl^2 + m^2} + m} } \right. 
 \nonumber \\
 && - \left. \frac{\pmi^2 {\bpms{}{+}}^2}
 {\pt{\esp{\pt{\varepsilon+\frac{1}{2}\omega+\mu}/T} + 1} J_1(\pmi R)^2 
 \pq{2R\pt{\pmi^2 + m^2} - \xi\sqrt{\pmi^2 + m^2} + m} }  \right\rbrace .
\end{eqnarray}
where we have defined $p_{\pm,\xi} = \frac{\zeta_{(\pm\frac{1}{2},\xi,l)}}{R}$
(see eq.~(\ref{zeta})).
We can rearrange the above sum by noting that the equation (\ref{zero}), depending on
indices $(M,\xi)$ is the {\em same} for $(-M,-\xi)$. In fact:
\begin{equation*}
 J_{\abs{-M-\frac{1}{2}}}(\zeta) \;+\; {\rm sgn}(-M)\,b^{(+)}_{-\xi}\,
 J_{\abs{-M+\frac{1}{2}}}(\zeta) = 
 \bsbd{\zeta} -\sgnm \bpms{-}{+}\bsbu{\zeta},
\end{equation*}
However, because of (\ref{bixi}), $-\bpms{-}{+}= \bpms{}{-} = 1/\bpms{}{+}$, and
so multiplying the right hand side of above equation by $\sgnm\bpms{}{+}$ one gets
the left hand side of eq.~(\ref{zero}). Hence, the zeroes of eq.~(\ref{zero}) and
the one with ``reflected" indices $(-M,-\xi)$ are the same:
\begin{equation}\label{parity}
   \zeta_{(-M,-\xi,l)} = \zeta_{(M,\xi,l)}
\end{equation}
for any $l=1,2,\ldots$. Now we can redefine the indices in the second term of the 
sum in eq.~(\ref{density3}) by turning $\xi$ into $-\xi$, which changes nothing 
as $\xi = -1,+1$ and write:
\begin{eqnarray*}
 D(0)^+
 &=& \frac{1}{8\pi^2 R} \sum_{\xi=\pm 1}\sum_{l=1}^\infty \int_{-\infty}^\infty 
 \di p_z \left\lbrace \frac{\ppl^2}{\pq{\esp{\pt{\varepsilon-\frac{1}{2}\omega+\mu}/T}
 +1} J_0(\ppl R)^2 \pq{2R\pt{\ppl^2 + m^2} + \xi\sqrt{\ppl^2 + m^2} + m} } \right. 
 \nonumber \\
 && - \left. \frac{\pmm^2 {\bpms{-}{+}}^2} {\pq{\esp{\pt{\varepsilon+\frac{1}{2}
 \omega+\mu}/T} + 1} J_1(\pmm R)^2 \pq{2R\pt{\pmm^2 + m^2} + \xi\sqrt{\pmm^2 + m^2} + 
 m} } \right\rbrace .
\end{eqnarray*}
We can replace $\pmm$ with $\ppl$ because of (\ref{parity}) and therefore:
\begin{eqnarray}\label{density4}
 D(0)^+ 
 &=& \frac{1}{8\pi^2 R} \sum_{\xi=\pm 1}\sum_{l=1}^\infty \int_{-\infty}^\infty 
 \di p_z \left\lbrace \frac{\ppl^2}{\pq{\esp{\pt{\varepsilon-\frac{1}{2}\omega +\mu}/T}+1} 
 J_0(\ppl R)^2 \pq{2R\pt{\ppl^2 + m^2} + \xi\sqrt{\ppl^2 + m^2} + m} } \right. 
 \nonumber \\
 && - \left. \frac{\ppl^2 {\bpms{-}{+}}^2} {\pq{\esp{\pt{\varepsilon+\frac{1}{2}
 \omega+\mu}/T}+1} J_1(\ppl R)^2 \pq{2R\pt{\ppl^2 + m^2} + \xi\sqrt{\ppl^2 + m^2} + 
 m} } \right\rbrace .
\end{eqnarray}

We are now going to prove that this latter expression is non-vanishing when $\omega 
\ne 0$. First, we note that it does vanish when $\omega=0$. In this case 
eq.~(\ref{density4}) yields:
\begin{equation*}
  D(0)^+\Big|_{\omega=0} = 
 \frac{1}{8\pi^2 R} \sum_{l,\xi} \int_{-\infty}^\infty \di p_z
 \frac{\ppl^2 \left[ J_1(\ppl R)^2 - {\bpms{-}{+}}^2 J_0(\ppl R)^2 \right]}
 {\pq{\esp{(\varepsilon+\mu)/T}+1} J_1(\ppl R)^2 J_0(\ppl R)^2 
 \pq{2R\pt{\ppl^2 + m^2} + \xi\sqrt{\ppl^2 + m^2} + m} } 
\end{equation*}
By using again (\ref{parity}) to replace $\ppl$ with $\pmm$ it is easy to show 
that the numerator of the integrand vanishes as:
\begin{equation*}
 J_1(\ppl R)^2 - {\bpms{-}{+}}^2 J_0(\ppl R)^2 = J_1(\pmm R)^2 - {\bpms{-}{+}}^2 
 J_0(\pmm R)^2 =  J_1(\zeta_{(-1/2,-\xi,l)})^2 - {\bpms{-}{+}}^2 
 J_0(\zeta_{(-1/2,-\xi,l)})^2 = 0
\end{equation*}
in view of the eq.~(\ref{zero}). Therefore, the spin tensor density in $r=0$ vanishes 
for a non-rotating system, as expected. To show that it is no longer zero for 
$\omega \ne 0$ we just need to show that the derivative with respect to $\omega/T$ 
in $\omega=0$ is not zero. One has:
\begin{equation*}
  \frac{\partial}{\partial (\omega/T)} D(0)^+\Big|_{\omega=0}
  = \frac{1}{16\pi^2 R} \sum_{l,\xi} \int_{-\infty}^\infty \di p_z
 \frac{\esp{(\varepsilon+\mu)/T} \ppl^2 \left[ J_1(\ppl R)^2 + {\bpms{-}{+}}^2 
 J_0(\ppl R)^2 \right]} {\pq{\esp{(\varepsilon+\mu)/T}+1}^2 J_1(\ppl R)^2 
 J_0(\ppl R)^2 \pq{2R\pt{\ppl^2 + m^2} + \xi\sqrt{\ppl^2 + m^2} + m} } 
\end{equation*}
All terms are mainfestly positive except $\pq{ 2R\pt{ p_{Tl}^2 + m^2} + 
\xi\sqrt{p_{Tl}^2 + m^2} + m }$ in the denominator when $\xi=-1$. However, in 
this case:
\begin{equation*}
 2R\pt{ p_{Tl}^2 + m^2} - \sqrt{p_{Tl}^2 + m^2} + m = \sqrt{p_{Tl}^2 + m^2} 
 \left( 2R\sqrt{p_{Tl}^2 + m^2}-1 \right) > \sqrt{p_{Tl}^2 + m^2} \, (2Rm-1)
\end{equation*}
which is positive for a radius greater than half the Compton wavelength of the 
particle, that is positive for any actually macroscopic value of the radius $R$.
The very same argument applies to the antiparticle term $D(0)^-$ of the $D(r)$
function in eq.~(\ref{density2}) with the immaterial replacement $\mu \to -\mu$,
hence:
\begin{equation*}
  D(0)|_{\omega=0} = 0  \qquad \qquad  
  \frac{\partial}{\partial (\omega/T)} D(0)\Big|_{\omega=0} > 0
\end{equation*}   

\end{document}